\documentclass[fleqn,usenatbib]{rasti}

\usepackage{newtxtext,newtxmath}
\usepackage[T1]{fontenc}

\DeclareRobustCommand{\VAN}[3]{#2}
\let\VANthebibliography\thebibliography
\def\thebibliography{\DeclareRobustCommand{\VAN}[3]{##3}\VANthebibliography}

\usepackage{graphicx}	% Including figure files
\usepackage{amsmath}	% Advanced maths commands
\usepackage{bm}        % Better bold math
\usepackage{algpseudocode}
\usepackage{algorithm}
\usepackage{enumitem}
\usepackage{orcidlink}
\newcommand{\bs}[1]{\ensuremath{\boldsymbol{\mathsf{#1}}}}

\def\matvis{\texttt{matvis}}
\def\pyuvsim{\texttt{pyuvsim}}

\title[Matrix-based visibility simulator for 21~cm arrays]{\matvis{}: A matrix-based visibility simulator for fast forward modelling of many-element 21~cm arrays}

\author[P. Kittiwisit et al.]{Piyanat  Kittiwisit\,\orcidlink{0000-0003-0953-313X}$^{1,2}$\thanks{Email: piyanat.kittiwisit@gmail.com},
	Steven G. Murray\,\orcidlink{0000-0003-3059-3823}$^{3,4}$,
	Hugh  Garsden\,\orcidlink{0009-0001-3949-9342}$^{5,6}$,
	Philip  Bull\,\orcidlink{0000-0001-5668-3101}$^{6,1}$,
\newauthor
	Michael J. Wilensky$^{6}$,
	Christopher Cain\,\orcidlink{0000-0001-9420-7384}$^{4}$,
	Aaron R. Parsons\,\orcidlink{0000-0002-5400-8097}$^{7}$,
	Jackson Sipple$^{7}$,
	Tyrone  Adams$^{2}$,
\newauthor
	James E. Aguirre\,\orcidlink{0000-0002-4810-666X}$^{8}$,
	Rushelle  Baartman$^{2}$,
	Adam P. Beardsley\,\orcidlink{0000-0001-9428-8233}$^{4,9}$,
	Lindsay M. Berkhout\,\orcidlink{0000-0002-2293-9639}$^{4,10}$,
\newauthor
	Gianni  Bernardi\,\orcidlink{0000-0002-0916-7443}$^{11,12,2}$,
	Tashalee S. Billings$^{8}$,
	Judd D. Bowman\,\orcidlink{0000-0002-8475-2036}$^{4}$,
	Richard F. Bradley$^{13}$,
\newauthor
	Jacob  Burba\,\orcidlink{0000-0002-8465-9341}$^{6}$,
	Steven  Carey$^{14}$,
	Chris L. Carilli\,\orcidlink{0000-0001-6647-3861}$^{15}$,
	Kai-Feng  Chen\,\orcidlink{0000-0002-3839-0230}$^{16,17}$,
	Samir  Choudhuri$^{5,18}$,
\newauthor
	Tyler  Cox$^{7}$,
	David R. DeBoer\,\orcidlink{0000-0003-3197-2294}$^{19}$,
	Eloy  de~Lera~Acedo$^{14}$,
	Matt  Dexter$^{19}$,
	Joshua S. Dillon\,\orcidlink{0000-0003-3336-9958}$^{7}$,
\newauthor
	Nico  Eksteen$^{2}$,
	John  Ely$^{14}$,
	Aaron  Ewall-Wice\,\orcidlink{0000-0002-0086-7363}$^{7,20}$,
	Nicolas  Fagnoni$^{14}$,
	Steven R. Furlanetto\,\orcidlink{0000-0002-0658-1243}$^{21}$,
\newauthor
	Kingsley  Gale-Sides$^{14}$,
	Bharat Kumar Gehlot$^{4}$,
	Brian  Glendenning$^{22}$,
	Adelie  Gorce$^{23}$,
	Deepthi  Gorthi\,\orcidlink{0000-0002-0829-167X}$^{7}$,
\newauthor
	Bradley  Greig\,\orcidlink{0000-0002-4085-2094}$^{24}$,
	Jasper  Grobbelaar$^{2}$,
	Ziyaad  Halday$^{2}$,
	Bryna J. Hazelton\,\orcidlink{0000-0001-7532-645X}$^{25,26}$,
\newauthor
	Jacqueline N. Hewitt\,\orcidlink{0000-0002-4117-570X}$^{16,17}$,
	Jack  Hickish$^{19}$,
	Daniel C. Jacobs\,\orcidlink{0000-0002-0917-2269}$^{4}$,
	Alec  Josaitis$^{14}$,
\newauthor
	Nicholas S. Kern\,\orcidlink{0000-0002-8211-1892}$^{16,17,\dagger}$,
	Joshua  Kerrigan\,\orcidlink{0000-0002-1876-272X}$^{27}$,
	Honggeun  Kim\,\orcidlink{0000-0001-5421-8927}$^{16,17}$,
	Matthew  Kolopanis\,\orcidlink{0000-0002-2950-2974}$^{4}$,
\newauthor
	Adam  Lanman$^{27}$,
	Paul  La~Plante\,\orcidlink{0000-0002-4693-0102}$^{7,8,28}$,
	Adrian  Liu\,\orcidlink{0000-0001-6876-0928}$^{7,29}$,
	Yin-Zhe  Ma\,\orcidlink{0000-0001-8108-0986}$^{30,31}$,
	David H.~E. MacMahon$^{19}$,
\newauthor
	Lourence  Malan$^{2}$,
	Cresshim  Malgas$^{2}$,
	Keith  Malgas$^{2}$,
	Bradley  Marero$^{2}$,
	Zachary E. Martinot$^{8}$,
\newauthor
	Lisa  McBride$^{23,29}$,
	Andrei  Mesinger\,\orcidlink{0000-0003-3374-1772}$^{3}$,
	Mathakane  Molewa$^{2}$,
	Miguel F. Morales\,\orcidlink{0000-0001-7694-4030}$^{25}$,
\newauthor
	Tshegofalang  Mosiane$^{2}$,
	Chuneeta Devi Nunhokee\,\orcidlink{0000-0002-5445-6586}$^{32,33}$,
	Hans  Nuwegeld$^{2}$,
	Robert  Pascua\,\orcidlink{0000-0003-0073-5528}$^{7,29}$,
\newauthor
	Yuxiang  Qin$^{34}$,
	Eleanor  Rath$^{16,17}$,
	Nima  Razavi-Ghods$^{14}$,
	James  Robnett$^{15}$,
	Mario G. Santos$^{2,1}$,
\newauthor
	Peter  Sims\,\orcidlink{0000-0002-2871-0413}$^{4}$,
	Saurabh  Singh\,\orcidlink{0000-0001-7755-902X}$^{29,35}$,
	Dara  Storer\,\orcidlink{0000-0003-4092-0103}$^{25}$,
	Hilton  Swarts$^{2}$,
	Jianrong  Tan$^{8}$,
\newauthor
	Nithyanandan  Thyagarajan\,\orcidlink{0000-0003-1602-7868}$^{36,15}$,
	Pieter  van~Wyngaarden$^{2}$,
	Zhilei  Xu\,\orcidlink{0000-0001-5112-2567}$^{16}$,
	Haoxuan  Zheng$^{17}$
\\
Affiliations are listed in Appendix~\ref{sec:affiliations}.
\vspace{-0.5cm}
}

\date{Accepted XXX. Received YYY; in original form ZZZ}

\pubyear{2025}

\begin{document}
\label{firstpage}
\pagerange{\pageref{firstpage}--\pageref{lastpage}}
\maketitle

% Abstract of the paper
\begin{abstract}
Detection of the faint 21 cm line emission from the Cosmic Dawn and Epoch of Reionization will require not only exquisite control over instrumental calibration and systematics to achieve the necessary dynamic range of observations but also validation of analysis techniques to demonstrate their statistical properties and signal loss characteristics. A key ingredient in achieving this is the ability to perform high-fidelity simulations of the kinds of data that are produced by the large, many-element, radio interferometric arrays that have been purpose-built for these studies. The large scale of these arrays presents a computational challenge, as one must simulate a detailed sky and instrumental model across many hundreds of frequency channels, thousands of time samples, and tens of thousands of baselines for arrays with hundreds of antennas. In this paper, we present a fast matrix-based method for simulating radio interferometric measurements (visibilities) at the necessary scale. We achieve this through judicious use of primary beam interpolation, fast approximations for coordinate transforms, and a vectorised outer product to expand per-antenna quantities to per-baseline visibilities, coupled with standard parallelisation techniques. We validate the results of this method, implemented in the publicly-available \matvis{} code, against a high-precision reference simulator, and explore its computational scaling on a variety of problems.
\end{abstract}

% Include between one and six keywords.

%%%%%%%%%%%%%%%%%%%%%%%%%%%%%%%%%%%%%%%%%%%%%%%%%%

%%%%%%%%%%%%%%%%% BODY OF PAPER %%%%%%%%%%%%%%%%%%

%=======================================
\section{Introduction} \label{sec:intro}
%=======================================
The 21~cm line emission from neutral hydrogen during the Cosmic Dawn and Epoch of Reionization (EoR) is currently one of the most challenging observational targets for radio astronomers. According to theoretical models \citep[e.g.][]{2018MNRAS.478.2193C}, the total line emission is expected to be quite weak, on the order of tens to hundreds of mK, with fluctuations of $\Delta_{\rm 21}^2 \lesssim 10^3~{\rm mK}^2$ (where $\Delta_{\rm 21}^2$ is the `dimensionless' power spectrum at a given scale and redshift). In comparison, the Galactic synchrotron foreground emission is many orders of magnitude brighter, with a much larger variance \citep{2005ApJ...625..575S}. Contributions from extragalactic sources are also considerably brighter than the target 21~cm signal. As such, 21~cm surveys must achieve both high sensitivity, to reach an adequate signal-to-noise ratio to detect the 21~cm fluctuations, and excellent instrumental stability and calibration precision, in order to prevent spurious leakage of some fraction of the bright foregrounds into the 21~cm signal modes. 

Even purpose-built 21~cm radio interferometer arrays have complex instrumental responses and systematics that interact with the sky and contribute to the difficulty of characterising the instrument sufficiently well to mitigate the foreground leakage problem mentioned above. These include the frequency- and direction-dependent polarised primary beam pattern of the antenna, and the time- and frequency-dependent receiver bandpass required to calibrate the data \citep{barry.etal.2016, barry.chokshi.2022}, the idiosyncratic properties of the receiver chain in the presence of radio frequency interference \citep[RFI;][]{wilensky.etal.2023}, environmental thermal variations \citep{gehlot.etal.2021}, reflection and retransmission of signals over the air and through cables \citep{kern.etal.2019,kern.etal.2020,josaitis.etal.2022}, among others.  In addition, the data analysis tools used to mitigate imperfections in the data (e.g. missing channels due to RFI contamination, filtering of cable reflections etc.) often have their own non-ideal behaviours, including the possibility of suppressing or filtering some of the 21~cm signal itself \citep{cheng.etal.2018,kolopanis.etal.2019,pagano.etal.2023}.

Physical (electronic and mechanical) approaches to instrumental characterisation may be useful, but also have their limitations in terms of understanding the ultimate impact of these systematics on the processed data. For example, primary beams can be characterised through a variety of physical methods, including photogrammetry, overflights by drones carrying standard sources and so on, but may still fail to capture the precise shapes of far sidelobes with sufficient accuracy \citep{jacobs.etal.2017}. Experimental approaches that isolate a single component of the system may also be unable to properly constrain the couplings and interactions between different components that give rise to some effects.

As a result, simulation-based, `forward-modelling,' approaches, for which realistic mock data is produced to study the complex process of observation and data analysis, are now a vital part of all 21~cm experiments. An important advantage of simulations is that they can be used to develop increasingly complex `what if' scenarios for the systematic effects that are identified, particularly for effects that would be close to (or even below) the noise level of the available data. Simulations can also be used as part of holistic efforts to `validate' the data analysis pipeline, allowing ones to explicitly determine levels of signal loss or other spurious features in the processing of the data as each step of the pipeline proceeds \citep{aguirre.etal.2022}. A disadvantage of simulation-based approaches is that they require models to be available for each system component, each systematic effect, and indeed the sources of emission on the sky, which can lack realism. They can also be computationally expensive, particularly when simulations for the whole array are required -- not only must we run the data analysis pipeline on a large volume of simulated data but we must also generate and store the simulated data itself.

The first step in the production of the mock data for the simulation-based study is the simulation of the complex visibility measured by radio interferometer arrays, and the basis of all of visibility simulators is a suitable simplification or special case of the radio interferometer measurement equation (RIME), which, as its name suggests, is the mathematical description of what an interferometer measures. 
Many visibility simulators have been developed, although only a small number of them are well documented. 
In general, they can be grouped into four categories: 
\begin{enumerate}[leftmargin=\parindent,align=left]
\itemsep0.5em
    \item Simulators that directly evaluate the original RIME formalism from \citet{hamaker.etal.1996, hamaker.2000, hamaker.2006}, which include simulators such as \texttt{OSKAR} \citep{mort2010oskar}, \texttt{WODEN} \citep{sullivan.etal.2012}, \texttt{healvis} \citep{lanman.kern.2019} and \pyuvsim{} \citep{lanman.etal.2019}.
    \item Simulators that evaluate the updated, generalised, RIME from \citet{smirnov.2011a,smirnov.2011b,smirnov.2011c,smirnov.2011d}, which can be used to describe all effects along the propagation path of the sky signals to the telescope, making the simulation results more suitable for testing direction-dependent calibration, for example. Such simulators include \texttt{MeqTrees} \citep{noordam.smirnov.2010} and \texttt{Montblanc} \citep{perkins.etal.2015}.
    \item  Simulators that utilise the `m-mode' spherical harmonic formalism in \citet{shaw.etal.2014,shaw.etal.2015}, which is more suitable for wide-field, transit, radio telescopes that use stationary antenna elements to capture the horizon-to-horizon sky signal as it transits across the arrays. Examples include \texttt{driftscan} \citep{driftscan}, and \texttt{RIMEz} \citep{martinot.2022}.
    \item General-purpose radio interferometric data analysis software with visibility simulation capability, including \texttt{FHD} \citep{sullivan.etal.2012,barry.etal.2019}, \texttt{CASA} \citep{thecasateam.etal.2022}, and \texttt{SAGECal} \citep{spreeuw.etal.2020a,spreeuw.etal.2020b,spreeuw.etal.2020c}. These tools usually simulate the visibility as part of the calibration model building step by evaluating either the original or generalised RIME in the Fourier transform domain.
\end{enumerate}  
Apart from the flavours of the RIME, all visibility simulators can vary significantly in realism, performance, and importantly accuracy.
The realism of the simulation is largely governed by the realism of the models used for each component of the RIME, such as the primary beam response, array layout and configuration, and sky intensity distribution.
Inclusion and implementation of specific models of these components by a simulator are largely dictated by its targeted applications and the experiments for which the simulators were originally developed. 
For example, \texttt{OSKAR} supports the use and customisation of beamformed stations, as it was developed as part of the Square Kilometre Array (SKA) study, while \texttt{WODEN} supports shapelet models of bright extended sources that are crucial for calibrating the the Murchison Widefield Array (MWA) 21~cm observations \citep{line.etal.2020}. 
Further instrumental and observational effects are usually added on top of the RIME as post-processing steps; there is no need to simulate RFI pulses through the visibility simulator machinery in most cases, for example. 

The performance of the simulators is governed by a few factors: the scale of the simulation, the particular choices of the algorithms and technology used to implement the RIME, and the approximations that are used in evaluating the RIME and its constituent component models. For instance, simulating `realistic' data volumes of a very large array, consisting of many antennas with wideband receivers, observing for a potentially very long period of time, can soon become prohibitively expensive, especially if multiple runs (e.g. to model different scenarios) are required. The performance of such a simulation can often be improved by employing parallelisation (e.g. \pyuvsim{}), vectorisation, or GPU acceleration (e.g. OSKAR, WODEN). 

One may also utilise approximations, such as analytical forms of the sky intensity and primary beam response that can be evaluated quickly, to improve performance at the cost of realism, but these approximations can also be important sources of inaccuracy. No simulator will be perfectly accurate, given limits to numerical precision and the uncertainty in the models of different components, but certain choices of algorithm and functional basis can give rise to inaccuracies that can potentially exceed the size of the effects that one is trying to model. Balancing realism, performance and accuracy can be complicated and depend on specific use cases as we have touched on. Regardless, it is important that the impact of the approximations and the overall accuracy of the simulator is well understood, validated, and documented for both users and other developers in the astronomical community, particularly when the simulation will be used to validate highly sensitive analyses of real data. 

In this paper, we introduce the \matvis{} visibility simulator and conduct rudimentary validation of its outputs. \matvis{} has been designed to perform fast simulation of the visibility response of many-element, drift-scanning, 21~cm arrays at low frequency, in particular the Hydrogen Epoch of Reionization Array \citep[HERA][]{deboer.etal.2017,berkhout.etal.2024}. The primary use case of \matvis{} outputs has been the validation of the HERA analysis pipeline via simulated mock data \citep{aguirre.etal.2022}. However, the underlying architecture of the simulator is more general and applicable to any radio interferometric array at any frequency. Similarly, its outputs can also be used in other applications with some limitations as we will discuss.
We elaborate on the design principles and use cases of \matvis{} in Section~\ref{sec:design}. In Section~\ref{sec:scheme}, we provide an overview of the basic calculation of a visibility simulator, as well as the specific algorithm and the approximations and assumptions used by \matvis{}. Section~\ref{sec:package} describes the software aspects of the publicly available \matvis{} package hosted on GitHub. In Section~\ref{sec:validation}, we perform a rudimentary `validation' of \matvis{} against \pyuvsim{}, an alternative simulator that has been designed to carefully and correctly evaluate each step of the RIME, at the cost of decreased computational performance. We explore effects of \matvis{} coordinate transformation scheme and show that \matvis{} can attain accuracy comparable to \pyuvsim{}. We also demonstrate the limitation of the point-like sky assumption in this section. Lastly, we study the computational scaling of \matvis{} to understand its performance in simulating ever-growing data sets on large arrays in Section~\ref{sec:scaling} and conclude in Section~\ref{sec:conclusions}.

%================================================
\section{Design Principles and Use Cases}\label{sec:design}
%================================================

The original motivation for \matvis{} was the need for a well-documented, well-tested, and fast visibility simulator to produce mock data for validating the analysis pipeline of the Hydrogen Epoch of Reionization Array \citep[HERA][]{deboer.etal.2017,berkhout.etal.2024}. This validation process has been deliberately described in \citet{aguirre.etal.2022}, for which its main goals of this process is to quantify potential signal loss from analysis steps through series of tests with mock data. In particular, pipeline validation requires production of mock data with `realistic physical properties' \textit{at the scale of the true data}. What constitutes these properties is an experiment-specific question, but generally includes reasonable scale-dependent magnitudes for different sky components (diffuse foregrounds, extra-galactic point sources and the 21~cm signal), with special attention to reproducing the spectral and temporal characteristics of the instrument.

This use case leads to the design principles for \matvis{} that naturally priorities the need for computational efficiency over detailed realism. In contrast, a use case of calibrating data would require exquisite realism and accuracy (both in the simulator and the input component models). To meet these requirements, \matvis{} introduces a matrix-based approach to evaluate the RIME, which is able to naturally take advantage of GPU acceleration. The primary novelty of \matvis{} is that it treats the RIME as an outer product over antennas, enabling the RIME for all baselines and polarisations at a single time and frequency to be computed as a single matrix product. 
The discretisation of the sky required to enable this matrix-based approach is predicated on being able to treat small, local equal-solid-angle, areas as delta-function point sources, essentially approximating the RIME angular integral as a Riemann sum, i.e. the `point-source approximation'. This treatment of the sky components has been elaborately studied and validated against analytical calculations in the \pyuvsim{} simulator \citep{lanman.etal.2019,lanman.etal.2022}, for which we compare and discuss in the paper. 
A further novelty (and approximation) of \matvis{} is that it utilises a coordinate transformation scheme that combines accurate-but-slow \texttt{astropy} \citep{astropy:2013, astropy:2018, astropy:2022} algorithms with simple-and-fast trigonometric-based calculations to significantly improve computational performance for simulations requiring multiple time-snapshots although this impact accuracy of the simulation.

These computational tricks employed by \matvis{} does not mean that it completely sacrifices accuracy. In fact, the accuracy of \matvis{} is equivalent to that of \texttt{pyuvsim} when simulating a single snapshot observation. Both simulators belongs to the first category of visibility simulators that we have described -- i.e. those that directly evaluate the RIME. Both assume the point-source approximation for sky models and use $E$-field primary beams (interpolated to given source positions and frequencies) in the calculations. The main differences are that \pyuvsim{} (i) directly calculates a full Jones-matrix contribution \citep{hamaker.2006} from each point source to each baseline, and (ii) uses \texttt{astropy} for coordinate transformation, providing the best possible source position accuracy. In contrast, \matvis{} uses (i) a matrix-based optimisation to improve speed of the calculations, and (ii) a geometric rotation scheme in place of \texttt{astropy} for coordinate transformation to trade some accuracy for more speed. The use of geometric rotations in \matvis{} does contribute to a lower accuracy than \texttt{pyuvsim} when simulating more than one time step at a time. We will demonstrate and discuss in Section~\ref{sec:validation}. \matvis{} is also similar to \pyuvsim{} in that it is more general in its applicability and this is reflected in the flexibility with which input components can be specified in the simulator. In terms of performance, however, we show in Section \ref{sec:scaling} that \matvis{} far outperforms \texttt{pyuvsim}. 

In context with other simulators that we have listed above, \matvis{} is likely in a similar performance regime as the GPU-accelerated \texttt{WODEN} and \texttt{OSKAR}. However, both are highly specific.  For example, \texttt{OSKAR} is built with simulating the SKA in mind, while \texttt{WODEN} has support only for built-in primary beams, which are limited to some analytic models and the MWA beams. Instead, \matvis{}, like \texttt{pyuvsim}, accept any external primary beam, and any sky model that can be represented as a list of spatially compact components.

Despite, there are a few limitation to the current version of \matvis{} (\texttt{1.2.1}) presented in this paper. First, it is currently not possible to simulate from fully polarised sky models although polarised beam is supported. This is due to the matrix-based algorithm that we will discuss shortly. Secondly, it currently only support simulating drift-scan observations. This limitation is the design choice specific to HERA and most of other 21~cm arrays, which are drift-scan instruments. Modification of our current algorithm to include supports for polarised sky and tracking observations is possible with some caveates. We will discuss them in Section~\ref{sec:scheme} .

%================================================
\section{Computational scheme} \label{sec:scheme}
%================================================

In this section, we introduce the algorithm used by \matvis{} to simulate visibilities. In common with all visibility simulators, a variety of choices need to be made in terms of how to model the sky (particularly whether and how to decompose it into a set of discrete elements); how to perform the integral over the whole sky required by the radio interferometer measurement equation (RIME); how to model the primary beams of the antenna elements; what coordinate systems to work in; and how to approximate or optimise computationally demanding steps of the process.

%-------------------------------------------------------
\subsection{Basic Framework} \label{sec:framework}
%-------------------------------------------------------

The purpose of the visibility simulation is to evaluate some form of the RIME \citep[e.g.][]{hamaker.etal.1996,smirnov.2011a} at the desired frequencies and observing times for a given sky model and instrumental model. Let $\boldsymbol{\Omega}$ be celestial coordinates for a point on the sky (e.g. equatorial coordinates in RA and DEC); then the RIME for a snapshot visibility (at a particular time $t$, and frequency $\nu$) can be written in a reasonably general form as \citep{kohn.etal.2019}
\begin{equation}
\begin{split}
    V_{ij}(\nu, t) = \int_{4\pi} \boldsymbol{\mathsf{A}}_i(\nu, t) \,\boldsymbol{\mathsf{C}}(\nu)\, &\boldsymbol{\mathsf{A}}_j^\dagger(\nu, t) \\
    &\exp \left ( -2\pi \nu \mathrm{i} \boldsymbol{b}_{ij} \boldsymbol{\cdot} \boldsymbol{n}/c \right ) \mathrm{d}^2\boldsymbol{\Omega}. 
\end{split}
\label{eq:rime}
\end{equation}
Here, $V_{ij}$ is the complex visibility for a baseline formed from antennas $i$ and $j$, and we have suppressed dependence on sky coordinate inside the $\boldsymbol{\mathsf{A}}$ and $\boldsymbol{\mathsf{C}}$ terms in the integral for notational clarity. In this expression, $\boldsymbol{\mathsf{A}}_i$ denotes the polarised primary ($E$-field) beam for antenna $i$, which is an $N_{\rm feed}\times N_{\rm ax}\equiv 2\times 2$ direction-dependent Jones matrix, whose components are the antenna patterns for each feed $p$, and two unit vectors needed to model the response to an incident $E$-field from spherical polar direction $k \in (\theta, \phi)$. Throughout, we will assume the case in which $p\in\{X,Y\}$, i.e. in which there are two perpendicular feeds, though \matvis{} trivially deals with the case in which there is a single feed. 
Note that $\boldsymbol{\mathsf{A}}$ is a special case of the full Jones matrix that determines the linear transformation of the electric field that propagates from the sky to the antenna. 
In the nomenclature of \citet{smirnov.2011a}, $\boldsymbol{\mathsf{A}}$ represents the full direction-dependent part of the Jones matrix (sans the phase difference term, which is captured explicitly by the exponential fringe factor in Eq. \ref{eq:rime}). 
In this paper, we consider $\boldsymbol{\mathsf{A}}$ to be fully specified by the primary beam of the instrument multiplied by a horizon mask. 
However, in principle it could include direction-dependent effects such as ionospheric refraction and Faraday rotation. 
While such effects are \textit{in principle} able to be included in simulations with \matvis{}, to do so would require manually altering the input primary beam model, and is not considered a natural use case for \matvis{} at this time. 
We will thus proceed to refer to $\boldsymbol{\mathsf{A}}$ as the `primary beam' within this paper.

Since we have chosen to represent the RIME as an integral over celestial coordinates, $\boldsymbol{\mathsf{A}}$ is represented as evolving over time, as the earth rotates.   
$\boldsymbol{\mathsf{C}}$ denotes the coherence matrix, which models the polarised flux distribution on the sky, which we \textit{assume} is static in celestial coordinates.\footnote{Of course, we could have represented the RIME in topocentric coordinates, in which case $\boldsymbol{\mathsf{A}}$ would have been static, and $\boldsymbol{\mathsf{C}}$ would evolve, but this representation is conceptually simpler for our algorithm presented later.} 
In terms of the Stokes parameters, $\boldsymbol{\mathsf{C}}$ is written as \citep{smirnov.2011a}
\begin{equation}
    \boldsymbol{\mathsf{C}} = \begin{pmatrix}
    \langle E_\theta E^*_\theta \rangle &  \langle E_\theta E^*_\phi \rangle \\
    \langle E^*_\theta E_\phi \rangle & \langle E_\phi E^*_\phi \rangle
    \end{pmatrix} = 
    \begin{pmatrix}
    I + Q & U + iV \\
    U - i V & I - Q
    \end{pmatrix}.
\end{equation}

Finally, the exponential term, which shapes the fringe pattern of the radio interferometer, includes a scalar product between the baseline vector $\boldsymbol{b}_{ij}$ connecting the antennas, and a unit vector $\boldsymbol{n}$ towards a particular point on the sky. The integral over solid angle is performed over $4\pi$ steradians, so that in principle the whole sky contributes to each complex visibility. In practice, only emission above the horizon at any particular time actually contributes, and this `horizon cutoff' can be bundled into the primary beam\footnote{The approach of bundling horizon effects into the beam is actually useful in terms of implementation because in practice the horizon may not always be perfectly flat \citep{bassett.etal.2021}, and therefore should be left flexible for a user to modify.}.

From here, all visibility simulators must make at least one approximation: we must make a choice about which discrete basis the sky model should be represented in, so that the integral in Equation~\ref{eq:rime} can be discretised. This is due to the fact that we do not have analytic forms of the sky or primary beam model as a function of sky coordinate. A very natural basis is the `point-source approximation', i.e. a choice of a set of points on the sky to which we assign the total flux density from their surrounding local region, and this is the representation that \matvis{} adopts. Note that while this form of discretisation is natural, it is by no means the only one. 
Discretisations in other kinds of bases exist, for example spherical harmonics \citep{shaw.etal.2014,shaw.etal.2015,martinot.2022}. 

Furthermore, \matvis{} makes a slightly more stringent point-source approximation for each sky pixel; it explicitly assumes that the full flux density from a given region defined as a `pixel' really does come from an infinitesimal point source located within the region, at which the primary beam and fringe term are evaluated. 
This point-source representation is in fact perfect if the sky purely consists of unresolved point-sources. However, if the sky contains extended sources, this approximation will inherently miss fluctuations on scales smaller than the sampling points (whether these fluctuations are intrinsic to the sky, or they come from the beam or fringe). In principle, this limitation can be overcome by following the Nyquist sampling principle, i.e providing input point sources dense enough on the sky that the smallest resolvable scales of the array are sampled \citep{lanman.etal.2022}.
We investigate the ramifications of this approximation on the accuracy of \matvis{} in detail in Section \ref{sec:diffuse}.

Summarily, \matvis{} uses a `point source' representation, treating the sky as a set of point sources with arbitrary frequency spectra binned into discrete frequency channels.
It does not enforce any particular `pixelisation', other than the condition that the sum of all pixels' flux densities equals the total flux density of the sky. A common example of such a pixelisation for diffuse emission is \textsc{HEALPix}\footnote{\url{http://healpix.sf.net/}} \citep{gorski.etal.2005}, but we emphasise that \matvis{} does not require a \textsc{HEALPix} representation, or even an equal-area pixel decomposition\footnote{Note that several well-used simulators adopt this same essential discretisation and associated approximation, including \texttt{pyuvsim} \citep{lanman.etal.2019},  \texttt{CASA} \citep{thecasateam.etal.2022} \texttt{miriad} \citep{Sault1995}. The latter two, which are more focused on narrow-field-of-view instruments and imaging, also adopt the specific pixelisation of a regular $(l,m)$ grid, and express their intensities in the commonly used units of Jy/pixel, which is only possible when all pixels have the same solid angle.}.  

From here on, we will refer to each component of the sky as a `source' or `point source' with the understanding that it may either be a truly physical point source, or an approximation of a region of solid angle as such. 
With this explicit assumption of a point-like sky model in place, we can write the RIME, at a particular frequency, as 
\begin{equation}
\begin{split}
    V_{ij}^{pq}(\nu_a, t) = \sum_{nkk'} \boldsymbol{A}^{pk}_{i}&(\nu_a,\boldsymbol{\theta}_{\mathrm{hrz}}^{(n)}(t)) \boldsymbol{\cdot} \boldsymbol{A}_j^{qk'*}(\nu_a,\boldsymbol{\theta}_{\mathrm{hrz}}^{(n)}(t)) \\
    & \times \boldsymbol{\mathsf{C}}^{(n)}_{kk'}(\nu_a,\boldsymbol{\theta}_{\mathrm{eq}}^{(n)})
    \exp\left(-2\pi \mathrm{i} \nu_a \tau^{(n)}_{ij}\right). 
\end{split}
\label{eq:matvisrime}
\end{equation}
This gives the complex visibility $V_{ij}^{pq}$ for antenna pair $i,j$ and polarisation $pq$. The sum is over all point sources in the sky model, and pairs of $E$-field axes.
$\boldsymbol{A}_i^{pk}$ is the complex primary beam response for antenna $i$ on feed $p$ to the $E$-field direction $k$.
The primary beam is evaluated at $\boldsymbol{\theta}_{\mathrm{hrz}}^{(n)}(t)$, the apparent angular position of each point source $n$ in a horizontal (topocentric) coordinate system at time $t$. The phase factor is a function of a baseline- and source position-dependent delay,
\begin{equation}
    \tau^{(n)}_{ij} = \boldsymbol{X}^{(n)}_{\rm hrz}(t) \boldsymbol{\cdot} \boldsymbol{b}_{ij} / c, \label{eq:matvisdelay}
\end{equation}
where $\boldsymbol{X}^{(n)}_{\rm hrz}(t)$ is the Cartesian unit vector pointing towards the source at an observed time $t$ in the same horizontal coordinate system as the baseline vector $\boldsymbol{b}_{ij}$ between the centres of antennas $i$ and $j$, and $c$ is the speed of light. 
We have chosen to write the source position unit vectors in the coordinate frames and representations that are more practical to evaluate for the different terms. We also note that the only approximation made in arriving at Equation~\ref{eq:matvisrime} has been the `point source approximation' that discretises the spatial integral.

%------------------------------------------------------------
\subsection{Matrix-Based Computation of the RIME} \label{sec:matrix_rime}

To obtain the visibilities, Equation~\ref{eq:matvisrime} must be calculated for all discrete frequencies, and at all requested discrete times, for all antenna pairs $i,j$. This becomes computationally prohibitive for large arrays. For reference, the completed HERA array will have 350 antennas (up to $\approx 61,000$ baselines), 1535 frequency channels, and so observing with a $\approx 10$~sec cadence over, say, 8 hours per night, we can see that Equation~\ref{eq:matvisrime} must be evaluated around $\sim 2.7 \times 10^{11}$ times to simulate even a single night of observing in full. Multiplying this by the number of individual sources in the sky model (e.g. the $\approx 300,000$ point sources in the GLEAM catalogue), this grows to $\sim 8 \times 10^{17}$. Clearly, the computational problem is large in scale.

A common approach to calculating Eq. \ref{eq:matvisrime} is to perform a summation over frequencies, times and \textit{baselines}, and for each, computing the primary beam product $A_i A^*_j$ and the fringe term $\exp(-2\pi i \nu_a \tau_{ij})$ for each source, before finally summing over the sources (this approach is taken, for example, by \pyuvsim{}).
This has two non-optimal ramifications for computational performance: first, naively this involves redundant interpolations of each antenna beam, since each antenna enters $\approx N_{\rm ants}$ times throughout the loop over baselines. 
This can be avoided, even within this common baseline-based approach, by \textit{first} computing all antenna-beam interpolations, and then drawing from these pre-computed values within the baseline loop. 
The second inefficiency in this approach is that the final sum over sources is a relatively small computational task (essentially a vector-vector inner product). It is well-known that modern computational algorithms for matrix-matrix multiplication are far more efficient than the equivalent loop over vector-vector inner products: for a square $N\times N$ matrix $M$, the computation of $P = MM$ via looping over all $N^2$ elements of $P$ and computing $P_{ij} = \sum_k M_{ik}M_{kj}$ scales as $N^3$, whereas modern matrix-matrix multiplication algorithms can scale as well as $\sim N^{2.3}$ \citep{williams.etal.2023,legall.urrutia.2017}.

The fundamental approach of the \matvis{} algorithm is to express Eq.~\ref{eq:matvisrime} as a tensor product (and, in some special cases, a matrix product) of tensors that are sized proportional to $N_{\rm ants}$ instead of $N_{\rm bls}$.
The tensor product `expands' the calculated quantities into a full set of cross-correlations for all baselines. 
This significantly reduces the number of repeated calculations, especially for a large array, and makes good use of the vectorisation capabilities of modern linear algebra frameworks and compute architectures (including GPUs). 
The downside is that all antenna-antenna correlations must be performed to make use of this approach, so it is not possible to be selective about which baselines are formed. For problems where the number of baselines to be simulated is close to the maximum for the array, $N_{\rm bls} \approx N_{\rm ants}(N_{\rm ants} - 1) / 2$, this approach will be highly efficient.

There are two main tricks required to implement this approach.
The first trick is to re-write the exponential fringe term as a product of antenna-based factors. Letting $\mathbf{y}_n = 2\pi \nu_a \mathbf{X}^{(n)}_{\rm hrz}(t)/c$ for compactness, we write
\begin{align}
    e^{ - \mathrm{i} \mathbf{y}_n(t) \mathbf{b}_{ij}} &= e^{-\mathrm{i} \mathbf{y}_n(t) \mathbf{x}_{i}} e^{+\mathrm{i} \mathbf{y}_n(t) \mathbf{x}_{j}} \\ &\equiv \mathsf{\mathbf{F}}_{in}(\nu_a, t)\mathsf{\mathbf{F}}^*_{jn}(\nu_a, t).
\end{align}
The second trick is to note that since the $2\times 2$ matrix $\boldsymbol{\mathsf{C}}_n$ for any particular source $n$ is hermitian, it can be written 
\begin{equation}
    \boldsymbol{\mathsf{C}}_n = \boldsymbol{M}_n\boldsymbol{M}_n^\dagger.
\end{equation}
We give the form of $\boldsymbol{M}$ in Appendix \ref{app:mmatrix}. With this in place, we can re-write Eq.~\ref{eq:matvisrime} as the tensor product
\begin{align}
    V^{pq}_{ij}(\nu_a, t) &= \sum_{k,k',n} \mathbf{A}_{ipkn} \mathbf{F}_{in} \mathsf{\mathbf{C}}_{kk'n}  \mathbf{F}^*_{jn} \mathbf{A}^*_{jqk'n}, \\
    &= \sum_{k,k',k''n} \mathbf{A}_{ipkn} \mathbf{F}_{in} \mathbf{M}_{kk''n}\mathbf{M}^*_{k'k''n}  \mathbf{F}^*_{jn} \mathbf{A}^*_{jqk'n},
    \label{eq:matvistensorprod}
\end{align}
where each of the factors in the sum is understood to be dependent on the frequency and time. Here $i$ and $j$ index antennas, $p$ and $q$ index the (up to) two feeds on each antenna, $k$ and $k'$ index the $E$-field component, and $n$ indexes the sky source/component.
Now, let 
\begin{equation}
    \boldsymbol{\mathsf{Z}}_{ip, nk''} = \sum_k \mathbf{A}_{ipkn} \mathbf{F}_{in} \mathbf{M}_{kk''n},
    \label{eq:fullz}
\end{equation}
be a \textit{matrix} (note the comma between $ip$ and $nk''$ in the indices, indicating that we are rolling two indices into one axis). 
Then, we simply have that
\begin{equation}
    V_{ij}^{pq}(\nu_a, t) = \boldsymbol{\mathsf{Z}}\boldsymbol{\mathsf{Z}}^\dagger,
    \label{eq:zz}
\end{equation}
i.e. the full visibility can be expressed as a matrix product of a $N_{\rm feed}N_{\rm ant} \times N_{\rm ax} N_{\rm src}$ matrix with its own conjugate transpose.
This matrix product produces the necessary two beam factors (with the Hermitian conjugate), squares the square root of the coherency matrix, and forms the necessary phase factor.

The required memory for this matrix product is quite reasonable, as there is simply one copy of an array that is of size $4 N_{\rm ant} N_{\rm source}$, and matrix-matrix products are extremely fast with modern algorithms and architectures, especially on the GPU.

We may go a little further if we assume that the sky is unpolarized.
In this case, the coherency matrix can be approximated as 
\begin{equation}
    \mathsf{\mathbf{C}}(\nu_a, t) = \begin{pmatrix}
        I(\nu_a) & 0 \\ 
        0 & I(\nu_a)
    \end{pmatrix},
\end{equation}
in which case we simply have
\begin{equation}
    \textbf{M} = \begin{pmatrix}
        \sqrt{I(\nu_a)} & 0 \\ 
        0 & \sqrt{I(\nu_a)}
    \end{pmatrix}.
\end{equation}
Having zero off-diagonal terms means that the sum over $k$ in Eq. \ref{eq:fullz} is no longer required, and we can simply write
\begin{align}
    \mathsf{\boldsymbol{Z}}_{ip,nk''} = \mathbf{A}_{ipkn} \mathbf{F}_{in} \sqrt{I_n},
    \label{eq:littlez}
\end{align}
which saves a rather large sum. The combination of Equation~\ref{eq:zz} and \ref{eq:littlez} is what \matvis{} currently implement in its algorithm.

There are a few important observations to be made about this final approximation. First, we again note that we have assumed that all sky sources are un-polarized.
This approximation has served well for the initial intended purpose of the \matvis{} software---producing simulations for the validation of the HERA analysis pipeline.
We note that even though the sky is approximated as unpolarized, the output visibilities still have non-zero cross-polarization, due to the inclusion of polarized beam sensitivity on each feed. 
Thus far, this unpolarized formulation is the only one implemented in the \matvis{} software (driven by these motivations), but it will be trivial to add full polarization support according to Eq.~\ref{eq:matvistensorprod} in future versions. 

Secondly, note that the peak memory usage for this algorithm is almost completely controlled by the number of antennas and feeds, as they are combined as an outer product. Both frequency and time are neither summed over nor interact with other frequencies and times, making them ideal as outer-loops for the algorithm. The source and field-component axes are summed over, and therefore may be `chunked' to suit any memory needs, at the expense of some vectorisation. Thirdly, we re-iterate that the formulation of Equation~\ref{eq:zz} requires that all baselines are simulated at once. This can be a disadvantage when simulating arrays with high number of redundancies although the optimisation implemented here still leads to several orders of magnitude speed up in comparison to a simulator that directly evaluates per-baseline visibilities in Equation~\ref{eq:matvisrime} such as \pyuvsim{} (see Section~\ref{sec:scaling}). Nevertheless, we believe that further optimisation to this algorithm is possible, and we describe the idea in Appendix~\ref{app:enhancement} to be fleshed out in future work.

%-------------------------------------------------------------
\subsection{Coordinate Transformations} \label{sec:coordinate}
%-------------------------------------------------------------
One of the key contributors to the accuracy of a visibility simulator, as well as its computational performance, is the manner in which the necessary coordinate transformations are carried out. On the one hand there is a sky model, which is most naturally represented in a celestial coordinate system, e.g. equatorial coordinates. On the other, there are the instrumental terms, particularly the primary beam response and the set of baseline vectors, which for stationary antennas like HERA, the MWA and SKA, are most naturally represented in a horizontal coordinate system connected to the location of the array. The complex exponential (fringe pattern) term in the RIME mixes the two, as it involves both the source position and baseline vectors. It is therefore always necessary to transform between celestial and horizontal coordinates at some point in the RIME calculation.

One way to perform these coordinate transforms is to use simplified trigonometric calculations that (i) converts from equatorial (RA and Dec, ($\alpha$ and $\delta$) coordinates to a Cartesian system via geometric vector projection, and then (ii) performs a solid body rotation into a horizontal coordinate system. The first step is,
\begin{equation}
    \boldsymbol{X}^{(n)}_{\rm eq} = \left ( \cos \alpha^{(n)} \cos \delta^{(n)},~~ \sin \alpha^{(n)} \cos \delta^{(n)},~~ \sin \delta^{(n)} \right ).
\label{eq:xeq}
\end{equation}
This vector of coordinates only needs to be calculated once for each source $n$ in the catalogue. The rotation matrix for (ii) is the same for all point sources for a given LST, and so there is also a computational saving from only needing to calculate it once per LST. Defining the horizontal coordinate system as an ENU (East-North-Up) topocentric system, we obtain the rotation matrix,
\begin{equation}
    \boldsymbol{\mathsf{R}}(t) = \left ( \begin{array}{ccc}
   \sin(H) & \cos(H) & 0 \\
   -\sin(l) \cos(H) & \sin(l) \sin(H) & \cos(l) \\
   \cos(l) \cos(H) & -\cos(l) \sin(H) & \sin(l)
\end{array} \right ), 
\label{eq:rotation}
\end{equation}
where $H$ is the hour angle of a reference point on the celestial sphere as a function of LST, and $l$ is the latitude of the array on the Earth. Then, the the horizontal coordinates of the sources are simply,
\begin{equation}
    \boldsymbol{X}^{(n)}_{\rm hrz}(t) = \boldsymbol{\mathsf{R}}(t) \cdot \boldsymbol{X}^{(n)}_{\rm eq}.
\end{equation}

Although this is fast, as it relies only on basic operations, and can be used to give a reasonable representation of the necessary coordinate transformations, it is also inaccurate owing to the lack of astrometric corrections. A number of time-dependent corrections for precession, aberration, and so on, are generally required to recover the apparent positions of celestial sources within arcsecond accuracy. In the case of \matvis{}, the fringe term, i.e. the exponent term in Equation~\ref{eq:littlez} which requires time-dependent source positions in horizontal coordinates, is particularly sensitive to positional errors, and we have found in our tests that the accuracy of the equatorial-to-horizontal coordinate transform is often the main driver of the precision of the simulator as a result. 

The easy solution is to use \texttt{astropy} as a full set of astrometric corrections has been implemented in its coordinate transformation module. However, these corrections can be expensive to compute, particularly for large catalogues containing many sources. During the early state of the development of \matvis{}, we have found that employing a full \texttt{astropy}-based coordinate transforms would result in the coordinate transformation step being the bottleneck, significantly limiting the performance of \matvis{}.\footnote{During the course of the publication of this paper, we have significantly generalised the implementation of sky-coordinate calculations within \matvis{}. In version from 1.3.0 onwards, the user is not restricted to using rigid rotations for the sky coordinates, but instead any coordinate transform may be used. In particular, the accurate transforms from \texttt{astropy} are built-in, along with new GPU implementations of the same, as well as schemes similar to the fast rigid-rotators with reference-time corrections described in this paper. The coordinate transformation schemes discussed in this paper are still applicable to version 1.2.1 and below, and have been used for all tests in Section 5. The details of the improvements mentioned here will be discussed in a future publication.} 

As such, we implement a two-step coordinate transformation scheme, which in our case are applied to the point sources, to find their locations in Cartesian horizontal coordinates at each LST. In short, we perform a pre-computation procedure to `perturb' the input equatorial coordinates of the sources so that steps (i) and (ii) in the trigonometric transforms (i.e. Equation~\ref{eq:xeq} and \ref{eq:rotation}) result in a more accurate observed source location when applied. These fictitious coordinates only need to be computed once, and significantly improve the accuracy of the results. Algorithm~\ref{alg:coordinate} details how we compute them.

\begin{algorithm}
\caption{Algorithm for \matvis{}'s `fast' coordinate transformation scheme.}
\label{alg:coordinate}
\begin{enumerate}[leftmargin=\parindent,align=left,itemsep=0.5em,label=(\roman*)]
    \item Given a reference time $t_{\mathrm{ref}}$, which can be different that the observation time of the simulation, compute the apparent local sidereal time at the location of the telescope via \texttt{astropy}.
    \item Compute $\boldsymbol{\mathsf{R}}^{\prime}(t_{\mathrm{ref}})$, the ENU-to-topocentric rotation matrix (inverse of Equation~\ref{eq:rotation}) at the reference time with the LST obtained from the above step and the latitude of the telescope.
    \item Convert the input equatorial source positions to horizontal coordinates in altitude and azimuth, $\boldsymbol{\theta}_{\mathrm{hrz}}$ through the \texttt{astropy}'s \texttt{SkyCoord.transform\_to} method, using $t_{\mathrm{ref}}$ as the observation time.
    \item Perform a geometric vector projection of $\boldsymbol{\theta}_{\mathrm{hrz}}$ into Cartesian coordinate to obtain $\boldsymbol{X}^{(n)}_{\rm hrz}(t_\mathrm{ref})$, essentially an equivalent of Equation~\ref{eq:xeq}. 
    \item Compute $\boldsymbol{X}_{\mathrm{eq,perturbed}}^{(n)} = \boldsymbol{\mathsf{R}}^{\prime}(t_{\mathrm{ref}}) \cdot \boldsymbol{X}^{(n)}_{\rm hrz}(t_{\mathrm{ref}})$
    \item Do the inverse of Equation~\ref{eq:xeq} to infer the RA and Dec from $\boldsymbol{X}_{\mathrm{eq,perturbed}}^{(n)}$. 
\end{enumerate}
\end{algorithm}

When steps (i) and (ii) are subsequently applied to the perturbed equatorial coordinates for other observation times in the simulation, we find a significant improvement to the accuracy of the observed source locations compared with the unperturbed ones. This correction degrades with increasing separation in time from the reference time although it will improve again once the separation is longer than $\approx12$ h, and reach the best improvement again at slightly less than 24 h from the reference time, because the correction is tied to the source positions. This also means that the source positions will matches those from the \texttt{astropy} transformation and thus errors from source will be minimum when simulating a snapshot observation at a single time. We show how the errors from this correction can affect our use cases of \matvis{} in Sect.~\ref{sec:validation}.

%-------------------------------------------------------
\subsection{Beam Interpolation} \label{sec:beam}
%-------------------------------------------------------
One of the most important factors in the accuracy of the simulator is how accurately the antenna beams are represented.  This is made more complicated by the fact that the beams must be evaluated at floating sky-coordinates defined by the discrete sky sources, as they drift in time. \matvis{} allows two main forms of antenna beam to be input: analytic beams, and simulated beams.  The former are easily evaluated at any coordinate (and frequency) by their nature, and typically in a very efficient manner. The latter presents more challenges; simulated beams are typically simulated on a discrete angular grid that cannot coincide with the floating sky-coordinates at which the beam must be evaluated. Therefore, they must be interpolated.

\matvis{} makes use of the \texttt{UVBeam} class from the \texttt{pyuvdata} package \citep{hazelton.etal.2017}, developed and maintained by the Radio Astronomy Software Group (RASG\footnote{\url{https://radioastronomysoftwaregroup.github.io/}}), to handle the beam information and interpolation of regularly gridded, HEALPix gridded, and \texttt{CST}-simulated beams. 
The \texttt{UVBeam} class also implements a number of interpolation methods (with more being added over time), which \matvis{} directly exploits in its CPU implementation. 
The order of interpolation in \texttt{UVBeam} is selectable, and can use either \texttt{numpy.RectBivariateSpline} or the \texttt{map\_coordinates} function from \texttt{scipy} internally. For example, this can be bilinear, bicubic, or quintic, on a regular grid of altitude and azimuth. 
Thus, while accurate beam interpolation is important for accurate simulations, \matvis{} out-sources this complexity, and arbitrarily accurate and/or performant interpolation schemes can be adopted.

This picture is not quite as simple for the GPU implementation. 
There are currently no GPU-accelerated beam interpolators within the \texttt{UVBeam} class, so beam interpolation on the GPU is implemented directly within \matvis{}. 
The simplest such interpolation is bilinear, and thus far this is the only scheme implemented for the GPU version of \matvis{}.

Bilinear interpolation is likely to be more approximate than bicubic or higher orders of interpolation, but we note this limitation is \textit{not} fundamental to the \matvis{} algorithm -- it is simply a practical limitation of the current version of the code. It will not be overly difficult to add higher-order interpolation schemes in the future.
In any case, the degradation in accuracy incurred by bilinear interpolation is strongly dependent on the inherent structure in the beam model (i.e. more structure at smaller scales will incur higher inaccuracy) and the resolution of the raw beam input. 
The latter is controlled by the user to some degree -- certainly for particular experiments, it is possible to re-simulate the beam with an EM solver (e.g. FEKO or CST) at higher resolution if required.
If this is not possible for practical reasons, the beam can simply be interpolated to a higher regular-resolution grid with higher-order interpolation before simulation with \matvis{}, and then the bilinear interpolation applied to that `upsampled' beam model. 
Thus, even the particular practical limitation of the current GPU incarnation of \matvis{} is not \textit{really} a limitation but can be overcome by preprocessing.

Regardless, for the particular purpose of validating the HERA analysis pipeline, which was the primary motivation for development of \matvis{}, we find that bilinear interpolation on our CST-modelled beam with resolution of 1 square degree \citep{fagnoni.etal.2021} is suitable as it does not hinder the power spectrum derived from the simulations. We will discuss this in Section~\ref{sec:validation}

One point to note concerning this interpolation is that it is essential that the regular coordinates of the simulated lattice be defined on a spherical grid which has the zenith as a pole. This is due to the fact that many common $E$-field beam responses have a discontinuity in their phase (recalling that the response is complex) for one of their components at zenith, which is properly interpolated only if the zenith is considered a pole and interpolated `away from'. 

We describe how the beam information (as well as other required inputs) are passed to \matvis{} in the next Section.

\subsection{Summary of Algorithm}
\label{sec:algorithm}

Having described the mathematical framework in the preceding sections, we now summarise the algorithm of \matvis{} as it stands in its present implementation\footnote{That is version 1.2.1 at the time of the publication of this article.}.

Firstly, we note that each step is to be performed at a single frequency. Thus, frequency forms our outer-most loop, and indeed it is envisaged that each frequency would typically be performed as a separate job on a high-performance computing scheduler, i.e. as an \textit{embarrassingly} parallel task. Our second, internal, loop is over times.

The following inputs are expected by the basic \matvis{} algorithm:
\begin{itemize}[leftmargin=\parindent,itemsep=0.5em,align=left]
    \item A set of antenna locations, $\boldsymbol{\mathsf{D}} = \{\boldsymbol{x}_1, \dots, \boldsymbol{x}_{N_{\rm ants}}\}$ in Cartesian East-North-Up (ENU) coordinates as an $N_{\rm ant} \times 3$ matrix.
    \item A beam model, $A_{\iota jk}(\nu, \boldsymbol{\theta})$ for each \textit{unique} antenna, $\iota$, feed $j$ and $E$-field component $k\in (\theta,\phi)$, that may be evaluated (or interpolated) to any particular set of topocentric coordinates. Note that two physically separate antennas that nevertheless are assumed to have the same beam need only be provided once.
    \item An integer-valued vector $a_i$ whose value for antenna $i$ is the index of its corresponding beam model, $\iota$.
    \item  A matrix of sky model source locations in Cartesian equatorial coordinates (ECI), $\boldsymbol{\mathsf{X}}_{\rm eq} = \{\boldsymbol{X}_{\rm eq}^{(0)}, \dots, \boldsymbol{X}_{\rm eq}^{(N_{\rm src})}\} \in \mathbb{R}^{3\times N_{\rm src}}$, where each source is given by Equation~\ref{eq:xeq}.
    \item A positive, real-valued vector of source intensities, $\boldsymbol{I} \in \mathbb{R}_{+}^{N_{\rm src}}$, where $\boldsymbol{I}_n = 2 \boldsymbol{\mathsf{C}}^{(n)}_{00} = 2\boldsymbol{\mathsf{C}}^{(n)}_{11}$.
    \item An array $\boldsymbol{\mathsf{R}}$ of shape $N_{\rm times}\times 3\times3$ containing the rotation-matrices (cf. Equation~\ref{eq:rotation}) required to convert from equatorial to topocentric coordinates at each time-step. These are computed \textit{outside} the main algorithm loop using any method the user desires (though utilities for computing these are provided in the package). These matrices also \textit{define} the times of the simulation (no other time information is given to the algorithm).
\end{itemize}

Then, for a particular frequency $\nu$ and time-index $t$, the \matvis{} algorithm follows Algorithm \ref{alg:matvis}.
\begin{algorithm}
\caption{Main algorithm for \matvis{}.}
\label{alg:matvis}
\begin{enumerate}[leftmargin=\parindent,align=left,itemsep=0.5em,label=(\roman*)]
    \item Rotate the pixels/sources above the horizon into topocentric frame: $\boldsymbol{\mathsf{X}}_{\rm hrz} = \boldsymbol{\mathsf{R}}_t \mathbf{X}_{\rm eq}$.
    \item Interpolate the beams onto source coordinates, i.e. produce the
       $N_{\rm beam}N_{\rm feed} \times N_{\rm ax} N_{\rm src}$ complex-valued matrix
       $\mathbf{A}_{\iota j, kl} = A_{ \iota jk}(\nu, \boldsymbol{X}_{\mathrm{hrz}}^{(l)})$ for unique antenna beam $\iota$, feed $j$, $E$-field component $k$ and source $l$.
    \item Compute the antenna-based exponent: $\boldsymbol{\mathsf{\tau}} = -2 \pi \mathrm{i} \nu \boldsymbol{\mathsf{D}} \mathbf{X}_{\rm hrz} / c$, which is an $N_{\rm ant}\times N_{\rm src}$ matrix.
    \item Compute the $N_{\rm ant}N_{\rm feed} \times N_{\rm ax}N_{\rm src}$ per-antenna complex-valued matrix $\boldsymbol{\mathsf{Z}}_{ij, kl} = \sqrt{\boldsymbol{I}}_l \boldsymbol{\mathsf{A}}_{a_i j, kl} \exp(\boldsymbol{\mathsf{\tau}}_{il})$, the matrix form of Equation~\ref{eq:littlez}.
    \item Compute the $ N_{\rm ant}N_{\rm feed} \times  N_{\rm ant}N_{\rm feed}$ visibility: $\boldsymbol{\mathsf{V}} = \boldsymbol{\mathsf{Z}} \boldsymbol{\mathsf{Z}}^\dagger$ (Equation~\ref{eq:zz}).
\end{enumerate}
\end{algorithm}
There are a few important details to note with respect to this algorithm. First, the present implementation of this algorithm only allows simulating drift-scanning observations, which is the common observing mode of 21~cm cosmology experiments such as HERA. This limitation is due to the fact that the algorithm simply evolves the input source matrix by making a horizontal cut and then computing the visibilities from all sources that are rising for each iteration in time. To enable simulating tracking observations, an additional step that adds time-dependent phase to each antenna would need to be implemented. This would have only a small performance impact, but is currently not in the scope of the \matvis{} development team.

Secondly, since steps (iv) and (v) are simply matrix products, highly efficient thread-optimised implementations exist to compute them on CPUs, eg. in BLAS\footnote{Basic Linear Algebra Subroutines: \url{https://www.netlib.org/blas/}.}. Furthermore, drop-in replacements exist to compute matrix products even more efficiently on GPUs, for example the \texttt{CUBLAS} package. This is highly advantageous, as step (vi) is the computation bottleneck for realistic modern large-$N$ arrays. 

Lastly, the beam interpolation in step (ii) need only be performed for the number of \textit{unique} antenna beams. In many cases of current interest, all antennas will be assumed to have the same beam, which makes this step negligible in terms of computation. Nevertheless, if the simulation is intended to capture beam irregularities across the antennas, this step can become highly non-negligible. However, GPUs are also extremely well-adapted to spatial interpolation; thus a well-tuned GPU implementation, e.g. by using texture mapping, can significantly reduce this cost.

%====================================================
\section{The {\matvis{}} Package} \label{sec:package}
%====================================================

The \matvis{} package defines an API and an algorithm for computing the RIME, with the intention that different implementations can be provided as drop-in replacements of each other.  Currently, the package provides two such implementations: \texttt{matvis.cpu} and \texttt{matvis.gpu}, which define CPU-based and GPU-based implementations respectively. Notably, the API computes the RIME at a single frequency only but includes a loop over time. The majority of the code (all of the CPU-based implementation) are written in pure Python, utilising the standard \texttt{numpy}, \texttt{scipy}, and \texttt{astropy} libraries. The GPU-based implementation replaces some of the \texttt{numpy} operations, particularly steps (ii)-(v) (as described in Section~\ref{sec:algorithm}), with CUDA11 code wrapped by \texttt{PyCUDA}. Each step of the algorithm is implemented as its own CUDA kernel, glued together in Python to maintain code simplicity. Notably, matrix products directly use the highly-optimised \texttt{cuBLAS} library. 

To maintain feature parity between the CPU-based and GPU-based implementations, the coordinate transformation scheme outlined in Section~\ref{sec:coordinate} is separately implemented in the function \texttt{equatorial\_to\_eci\_coords} in the \texttt{conversions} module of the Python package because a GPU-based \texttt{astropy} coordinate transformation module does not exist. Users can manually call this function to put sources into the `perturbed' equatorial coordinates (as described in Section~\ref{sec:coordinate}) before passing them to the main \matvis{} routine to improve accuracy of the coordinate rotation in step (i). Although this is not required, i.e. the API has no knowledge of whether this correction has been applied, we highly advise that all simulations use it (see Section~\ref{sec:validation} on how this influence the simulation outputs).

As discussed, interfacing and interpolation of gridded beams are handled by the \texttt{UVBeam} from the \texttt{pyuvdata} package developed and maintained by RASG. For analytic beams, uniform, Gaussian, and Airy beams are readily supported through the \texttt{AnalyticBeam} class of the \pyuvsim{} package, which is also developed and maintained by RASG, and shares a common beam evaluation interface with the \texttt{UVBeam} interpolation. Other forms of analytic beams are possible by sub-classing the \texttt{AnalyticBeam} class, one of which that we have developed is the \texttt{PolyBeam} class in the \texttt{hera\_sim} package, which implements an analytical model of the HERA Phase I primary beam \citep[][see Section~\ref{sec:package}]{choudhuri.etal.2021}. 

A high-level wrapper is available via the \texttt{simulate\_vis} routine, which provides a loop over frequencies, as well as simple implementations to compute the required input equatorial sky coordinates, $\mathbf{X}_{\rm eq}$ from the more common RA/DEC. This high-level wrapper also provides a simple switch between different back-end implementations (i.e. CPU and GPU). We note that this wrapper does \textit{not} provide in-built capability to perform the coordinate perturbations (though pre-perturbed coordinates can be computed and provided to the wrapper by using the above mentioned function).
 
A separate wrapper is also available via the \texttt{visibilities} module in the \texttt{hera\_sim} package, a simulation package for HERA-like redundant interferometric arrays, also developed by the HERA collaboration. This has a few key advantages over the in-built wrapper in the \matvis{} package, including an option to apply coordinate correction given a reference time, a unified simulation interface that can be used to wrap other simulator codes (with \pyuvsim{} wrapper already implemented), and integration with systematics models available in the \texttt{hera\_sim} package. The main script of this wrapper, \texttt{hera-sim-vis.py}, uses \pyuvsim{}-style configuration files\footnote{See \url{https://pyuvsim.readthedocs.io/en/latest/parameter_files.html}}, making it much easier to configure and execute a large set of simulations on an HPC cluster. The \texttt{hera\_sim} package also implements an analytical model of the Phase I HERA dipole beam \citep{choudhuri.etal.2021} in its \texttt{PolyBeam} class for use with \matvis{} (and \pyuvsim{}) although we note hat the \pyuvsim{}-style configuration files do not currently support this beam (or any analytic beam models implemented outside of the \texttt{AnalyticBeam} class of the \pyuvsim{} package), and thus users must manually build \matvis{} simulations via the \texttt{hera\_sim} unified interface to use this beam. 

The \matvis{} package\footnote{\url{https://github.com/HERA-Team/matvis}} and all of its associated software that we have mentioned in this paper, including \texttt{hera\_sim}\footnote{\url{https://github.com/HERA-Team/hera_sim}}, \pyuvsim{}\footnote{\url{https://github.com/RadioAstronomySoftwareGroup/pyuvsim}}, and \texttt{pyuvdata}\footnote{\url{https://github.com/RadioAstronomySoftwareGroup/pyuvdata}}, are publicly available via open-source licenses on \texttt{GitHub}. PyPI distributions are also available, allowing easy installation via the \verb|pip install| command. \matvis{} has been under active development since May 2021. The development cycle employs \texttt{git} and \texttt{GitHub} and strictly follows an open-source development guideline laid out by the HERA collaboration\footnote{\url{http://reionization.org/wp-content/uploads/2013/03/HERA067_HERA_Software_Community_Guide_6May2019.pdf}}. 
Particularly, we adopt continuous development and comprehensive code review processes, thorough unit-testing, including for \pyuvsim{} and \texttt{pyuvdata}, and maintain detailed and user-friendly documentation\footnote{\url{https://matvis.readthedocs.io}}, including both API-level `docstrings' and a set of tutorials. We use Continuous integration (CI) to run tests and report coverage. 
Although the primary contributions to the \matvis{} code have thus far come from members of the HERA collaboration, we welcome and appreciate contributions from outside of the collaboration.

%==========================================
\section{Validation} \label{sec:validation}
%==========================================
In this section, we perform several tests to demonstrate suitability of \matvis{} simulations for validation of 21~cm power spectrum analyses. 
There are two important metrics when quantifying the accuracy of mock visibilities, and thus the visibility simulators, for a 21~cm power spectrum analysis. 
The first is that the dynamic range of the delay spectrum of foreground visibilities must be $>10^{10}$ to allow discerning the foreground and 21~cm signals at high delay. 
Secondly, the simulated 21\,cm visibilities must yield estimated power spectra that accurately match the input 21\,cm power spectrum.  
We will show that \matvis{} meets these two criteria in the following two subsections. 
Throughout these tests, we will also compare the accuracy of \matvis{} against a reference implementation of the RIME in the \pyuvsim{} simulator, which we discussed in Section~\ref{sec:design}. 

In addition to testing the simulator outputs on the targeted pipeline validation use case, we elaborate on the impact of the point-source approximate used by \matvis{} (as well as \pyuvsim{} and some other simulators), particularly with regard to the discretisation of the diffuse sky emission. We will provide a demonstration in the context of the delay spectrum analysis. We also touch on the imaging of simulated visibilities from \matvis{}, with a focus on the impact of \matvis{}' `fast' coordinate transformation scheme on the point source position in the image.

With these tests, we aim to demonstrate the accuracy that \matvis{} can achieve on the type of experiments that it is designed to simulate and for the targeted scientific measurements that those experiments aim to measure. However, we also note that the results presented here should not be treated as an absolute measure of its accuracy. The numbers will change depending on the simulation parameters and sky models, and whether they are sufficiently accurate will depend on specific use cases.

The following software versions are used throughout these tests: \texttt{matvis==1.2.1}, \texttt{hera\_sim==4.2.0}, \texttt{pyuvsim==1.2.6}, and \texttt{pyuvdata==2.4.0}. For all simulations, we use a the CST-simulated HERA beam from \citet{fagnoni.etal.2021} with bi-linear and bi-cubic interpolations in spatial and frequency dimensions respectively.

\subsection{Delay Spectrum Dynamic Range} \label{sec:dps}

One of the key concerns in 21~cm power spectrum analysis is the dynamic range of the foreground signal. This is especially important for the foreground-avoidance delay-spectrum technique used by HERA \citep{parsons.etal.2012}. The basis of this measurement is that the clean and calibrated visibility is Fourier transformed in frequency with a chosen taper function before forming a power spectrum over the delay modes. The foreground power will be smooth and concentrate around zero delay, forming a peak, and taper off at high delay, if there are no other systematics. The 21~cm delay spectrum, on the other hand, will be relatively flat, unsmooth, and lies $\sim4-5$ orders of magnitude lower than the foreground power at zero delay. 
Thus, the dynamic range between the peak and the floor of the foreground delay spectrum is an important factor to allow discerning of the foreground and 21~cm signals.

To demonstrate that \matvis{} is able to produce simulations that yield delay spectra of sufficient dynamic range, we perform simulations with the same sky models used in the validation of the recent 21~cm power spectrum upper limits from \citet{hera.2023}, and derive delay spectra from the simulation outputs. 
These sky models consists of a diffuse Galactic synchrotron emission model taken from the Global Sky Model \citep{gsm.2008} represented as a \texttt{HEALPix} map with $N_{\rm side}=256$, a full-sky mock point source catalogue derived from source count distribution of the GLEAM catalogue \citep{franzen.etal.2019}, and the extremely bright `A-team` sources extracted from the GLEAM catalogue \citep{hurley-walker.etal.2017}. We perform the simulations at two times, LST $=22.75$ hour (Field A) and LST $=5.13$ hour (Field C), each with 5-second integration over two frequency bands, 117.09 - 132.62 MHz (Band 1) and 150.29 - 167.77 MHz (Band 2). These fields and frequency bands correspond to those used for measurements in \citet{hera.2023}. We simulate two different baseline lengths in East-West and North-South direction, which correspond to the shortest and longest baselines in those directions in the current HERA array. We calculate the delay spectrum as discussed above with a Blackman-Harris taper function. Source position correction is applied as discussed in Section~\ref{sec:coordinate}, for which we use the simulated times as the reference times for coordinate transformations. This is essentially equivalent to no approximation on the source positions.

\begin{figure*}
    \centering
    \includegraphics{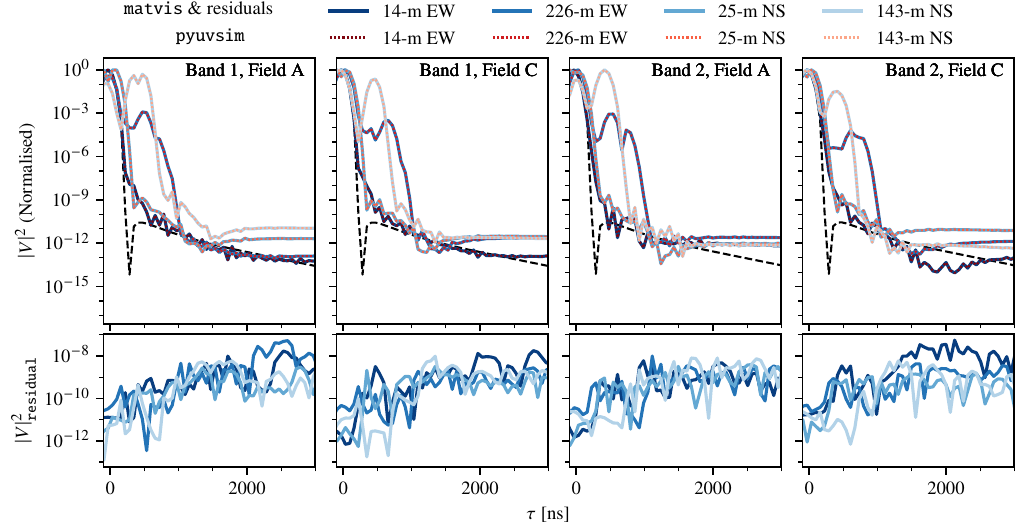}
    \caption{Delay spectra from \matvis{} and \pyuvsim{} simulations with mock foreground models at two different frequency bands and times at different baseline lengths and directions (top), and the fractional residuals between the results from the two simulators (bottom). The dashed line shows the delay spectrum of the Blackman-Harris taper function, which indicates the maximum possible dynamic range. All power spectra are peak-normalised to unity to emphasise their dynamic range. \matvis{} simulations yield delay spectra with dynamic ranges of $>10^{10}$ in all cases, and agree with \pyuvsim{} to better than $10^{-10}$\%.}
    \label{fig:dps}
\end{figure*}

Figure~\ref{fig:dps} shows the results from this test. The top row shows delay spectra derived from both simulators at different baseline lengths and directions within the core of the HERA array. The dashed line shows delay  spectra of the Blackman-Harris taper function, providing a baseline of how well one can possibly do in terms of dynamic range. 
We also peak-normalised all power spectra in the top row to make it easier to compare their dynamic ranges. The bottom row shows the fractional residuals between \matvis{} and \pyuvsim{}, $|V|_{\mathrm{residual}}^2=\left||V|_{\mathrm{matvis}}^2/|V|_{\mathrm{pyuvsim}}^2-1\right|$.  \matvis{} produces delay spectra with dynamic ranges of more than 10 orders of magnitude in all cases; 
the cases with the lowest dynamic range are the longest baselines, which likely suffer some aliasing of the fringe from the diffuse sky model (indicating a deficiency in the input model, rather than the simulator itself).
This interpretation of a sky-model deficiency is supported by the exceptional agreement of \matvis{} and \pyuvsim{}---at better than a millionth of a percent---for all times, frequencies and baseline lengths. 
Given the extensive testing of \pyuvsim{} (including against analytic solutions \citep{lanman.etal.2022}), this suggests that--beyond simply supporting power-spectrum validation efforts--\matvis{} could also be used for detailed calibration purposes, so long as the input sky models are sufficiently detailed, since the rule-of-thumb precision required for such purposes is about $10^{-10}$ in power-spectrum units \citep{barry.etal.2016}.

\begin{figure*}
    \centering
    \includegraphics{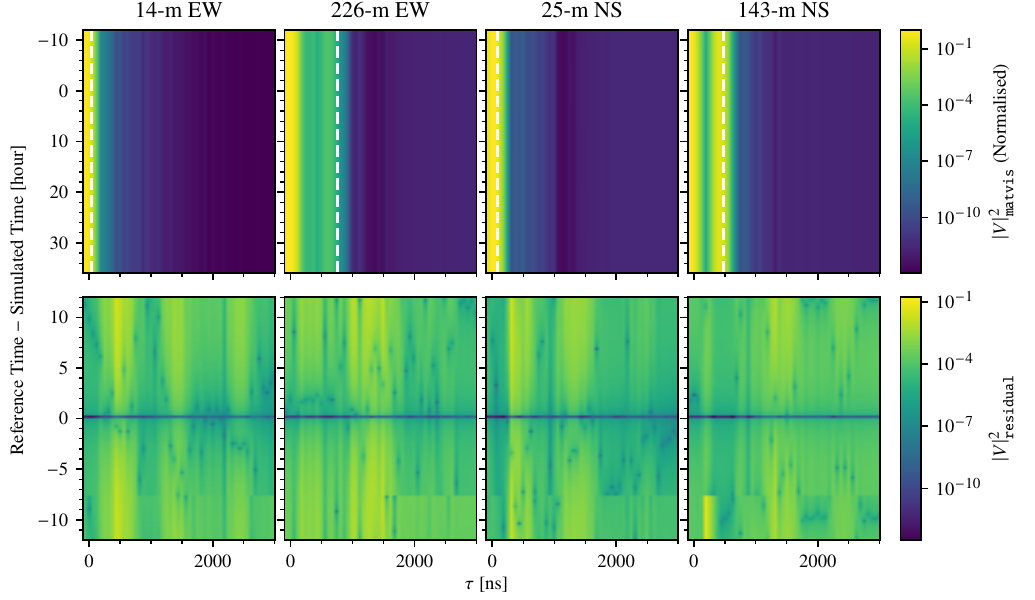}
    \caption{Foreground delay spectra from \matvis{} simulations with different coordinate transformation reference time (top) and fractional residuals against \pyuvsim{} (bottom). It is evident from the residuals that setting the reference time further away from the simulated observation time will results in errors, the larger the differences, the higher the errors. If accuracy is required, e.g. using \matvis{} simulation for calibration time, we recommend splitting the simulation into $<2$-hour time chunk, or simulate one time step at a time for best results. This errors, however, do not affect the delay spectrum dynamic range as also evident in the top row.}
    \label{fig:dps_ref_time}
\end{figure*}

Agreements with \pyuvsim{} will degrade if the reference time used for the coordinate transformation differs from the simulated time, for reasons that we discussed in Section~\ref{sec:coordinate}. 
Figure~\ref{fig:dps_ref_time} demonstrates this point. Here, the top row shows a heat map of delay spectra from \matvis{} simulations of field A in frequency band 1. In these maps, the vertical axis captures the offset between the reference and simulation times, and we simulate every 15\,minutes. The bottom row shows their fractional residuals against \pyuvsim{}. 
As expected, \matvis{} agrees best with \pyuvsim{} when simulating directly at the reference time of the coordinate transformation. When the simulated time differs from the reference time, discrepancies between the two simulators increase rather quickly, with $|V|^2_{\mathrm{residual}}$ doubling when the reference time is $\approx \pm1$ hour from the simulated time. However, the dynamic ranges of the power spectra do not degrade as the simulations move away from the reference time, as illustrated by the vertical constancy of the maps in the top panel.
Here, the dashed white lines indicate the horizon delay \citep{parsons.etal.2012}, and we note that at all times, the delay spectra achieve the required 10-12 orders of magnitude of dynamic range at or around the horizon.  
Thus, we consider the simulations sufficient for the purposes of validating power-spectrum analyses (for which the dynamic range and overall magnitude are the foremost concerns), but too discrepant at large offsets from the reference time for the purposes of calibration.
For the latter, we suggest splitting the simulation into chunks of 15 minutes or less, and performing the coordinate correction (c.f. \S\ref{sec:coordinate}) at a central reference time for each chunk.

\subsection{21 cm Power Spectrum}\label{sec:eor_ps}

The ability to obtain an accurate power spectrum estimate from the visibility simulated from a known 21~cm model is also an important factor that goes hand-in-hand with the delay spectrum dynamic range for a visibility simulator to be feasible for 21~cm power spectrum analysis. We test this by simulating from a 21-cm EoR analogue model similar to the one used in \citet{hera.2023}. The model is a Gaussian random field with a power spectrum slope, $P(k) \propto k^{-2.7}$, created using the \texttt{gaussian\_random\_fields}\footnote{https://github.com/zacharymartinot/redshifted\_gaussian\_fields} code. The EoR Gaussian field is drawn at pixel locations of a set of HEALPix maps with NSIDE=1024, each map corresponding to a different redshift and simulation frequency. We then discretise this set of maps into an ensemble of point sources as required by \matvis{} by multiplying the intensity of each pixel (in Jy/sr) by the pixel area, converting each pixel into a `point source' with flux density in Jy.

In general, we expect realistic sky models to be strictly positive-valued, because foregrounds are the dominant component. 
However, the 21\,cm signal is a differential quantity, and can be measured in absorption or emission. 
When simulating the 21\,cm signal in isolation from the foregrounds, we thus expect to encounter negative values, which are a problem for the \matvis{} simulator as it is currently implemented. 
To deal with the negative values for this test, we add a constant value to the entire 21\,cm sky model (all pixels at all frequencies) such that it becomes strictly positive.
This is equivalent to adding a monopole and will result in a bump in the power spectrum at low $k$, but should not modify the estimated power at larger $k$. 
This monopole bump can be removed by simulating just the monopole component separately and subtracting the resulting visibilities from the monopole-added visibilities. 

For the purposes of this test, we run \matvis{} simulations using a small HERA-like configuration: 19 antennas (hexagonal arrangement with 3 antennas on each side), 1 snapshot integration at LST = 0 hour, and 131 frequency channels spanning the Band 1 frequencies (117.09 -- 132.62 MHz). 

\begin{figure*}
    \centering
    \includegraphics[width=\textwidth]{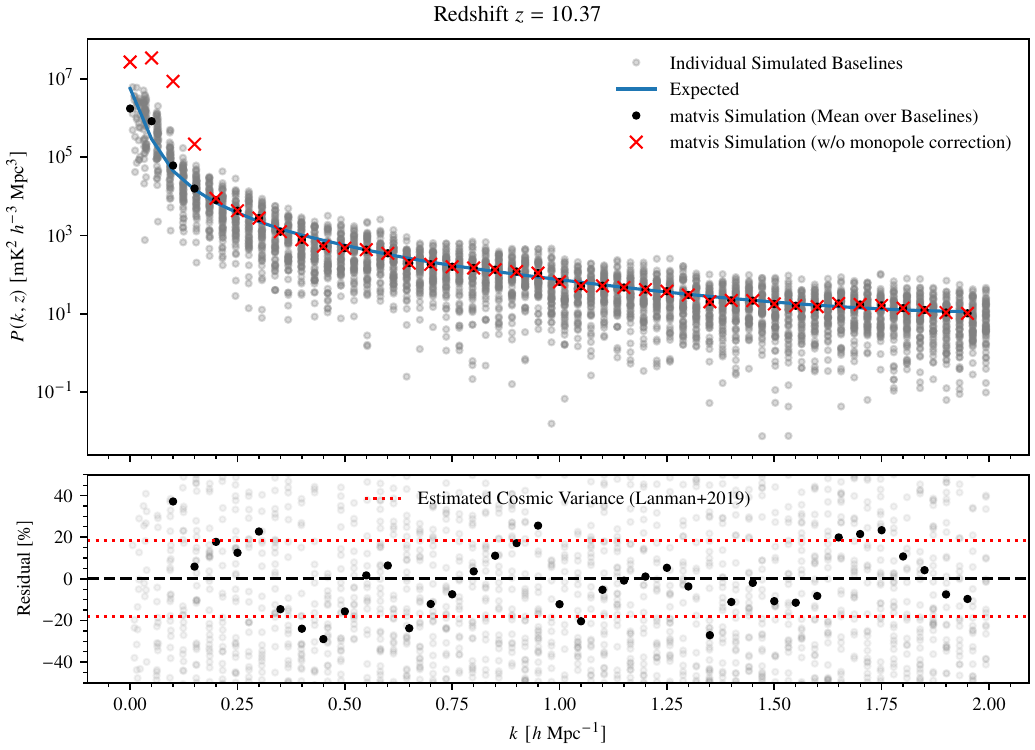}
    \caption{21~cm power spectrum estimates from \matvis{} simulations. The blue lines show the expected power spectrum of input model used in the simulation. This model has some negative values, and thus the a monopole has been added by shifting values to be all positive to conform with \matvis{} formalism. The power spectrum estimates from the resulting visibility shown in red crosses exhibit a monopole bump at low $k$. This bump is corrected by separately simulating the monopole and subtracting it before estimating the power spectrum. The estimates with this correction are shown in black circles, which follows the expected power spectrum at all $k$. Residuals between the expected and estimated power spectra are primarily due to cosmic variance, for which we show the level predicted by \citet{Lanman2019.cosmicvar} for HERA as the red dotted lines in the lower panel.}
    \label{fig:eor_ps}
\end{figure*}

We form the power spectra with the \texttt{hera\_psepc}\footnote{https://github.com/HERA-Team/hera\_pspec} code. 
This estimate is a complicated matter, and we suggest interested readers to consult, e.g., \citet{hera.2023} or \citet{gorgc.etal.2023} for details.
The top panel of Figure~\ref{fig:eor_ps} shows the resulting power spectrum estimates. The input power spectrum, corrected for expected aliasing and instrumental window-function effects \citep{aguirre.etal.2022}, is shown in blue. The red crosses are the baseline-averaged power spectrum from the \matvis{} visibility without monopole correction, which exhibit a prominent bump at $k < 0.20$. Although this bump seems problematic, it falls beneath the foreground power in those $k$ modes, and thus will not be an issue if the visibility is used for validating foreground avoidance analysis used by HERA. 
Regardless, it is simple to correct for the monopole (as discussed above), and the estimates with this monopole correction are shown as black circles. These follow the expected power spectrum well for all $k$. Individual-baseline estimates with monopole corrections are shown as grey circles for completeness.

Discrepancies between the estimates and the input power spectrum are expected to arise simply due to sample (sometimes also called `cosmic') variance. 
The expected sample variance is a decreasing function both of the total integration time and the number of unique baselines at each $k$ \citep{Lanman2019.cosmicvar}. 
We use the sample variance prediction from \citet{Lanman2019.cosmicvar} as integration time approaches zero, calculated for the HERA beam (their figure 5), to estimate the sample variance we expect. 
For a single baseline, the expected RMS fluctuations from smaple variance are 100\%, and we divide by $\sqrt{N_{\rm bl}} \approx 5.5$, yielding 18\% RMS (red dotted line in bottom panel of Fig. \ref{fig:eor_ps}). 
The residuals of the estimated power spectra to the input are generally within the expected sample variance.

\subsection{Simulating from Diffuse Sources}\label{sec:diffuse}

As discussed extensively already, \matvis{} assumes a point-source approximation for a sky model. To simulate visibilities from a resolved radio galaxy with extended jets or diffuse emission, which include the 21~cm signal, the user must first discretise the sky intensity into point sources, essentially a collection of delta functions that are assigned the integrated flux density of surrounding regions. 
\matvis{} is agnostic to the specific choice of of this discretisation, i.e. the location of each point source and the shape of the region surrounding it, within the dual limitations that (i) the sum of all discretised sources should yield the integrated flux density of the full sky, i.e. the discretisation should be a complete tessellation of the sky, and (ii) the fidelity of the simulation will in general be higher when the pixels are more compact (e.g. long `stripes' as pixels will be a poorer discretisation---for the same number of sources---than pixels with small `circular` surrounding regions).
For diffuse emission, one choice of explicit discretisation that is consistent with these assumptions is the HEALPix coordinate grid, by which each of the HEALPix pixel is treated as a point source. However, the calculation of the RIME is expected to break down on long baselines for a lower resolution HEALPix grid because the fringe terms (the exponent in the RIME), which depends on the source positions and baseline lengths, would not be evaluated enough to sufficiently sample the fine fringe patterns of the long baselines.

\begin{figure*}
    \centering
    \includegraphics[width=\textwidth]{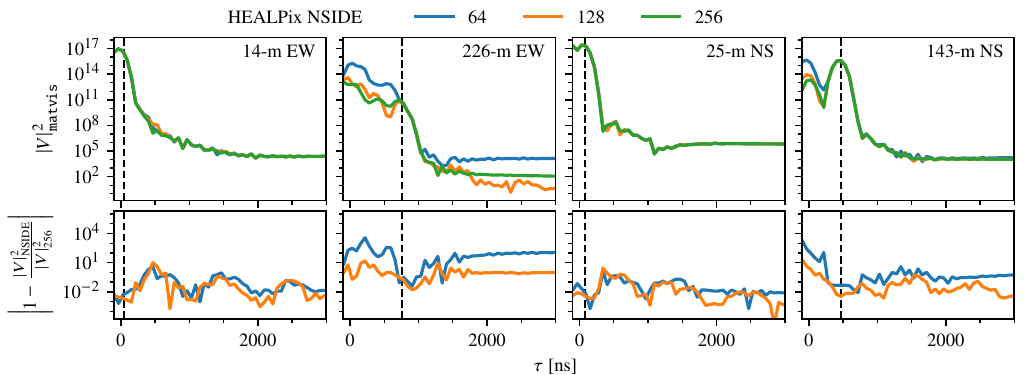}
    \caption{Demonstration of the effects of insufficient resolution in the discretisation of a diffuse sky model on the delay spectrum. We simulate a diffuse sky model discretised onto a HEALPix grid at different resolutions defined by the NSIDE parameter, and then form the delay spectra from the simulated output. At a low sky model resolution, the fringe term, which depends on the source positions, may not be evaluated sufficiently for long baselines (e.g. second and fourth panels) with fringe spacing shorter than the approximate angular resolution of the discretization. This aliasing results in errors in the visibility that propagate to the delay spectrum.}
    \label{fig:diffuse_ps}
\end{figure*}

To demonstrate, we compare the delay spectra of simulation outputs of solely the diffuse component for Band 1 (117.09 -- 132.62 MHz) and Field C from Section~\ref{sec:dps}, which was discretised on a HEALPix grid with NSIDE=256, to those derived from simulations from the same diffuse models discretised on lower resolution HEALPix grids. Figure~\ref{fig:diffuse_ps} shows these results, for which the top row is the derived delay spectra, and the bottom row is the fractional residuals between the spectra derived from a lower resolution sky model to that of the NSIDE=256 model. The NSIDE parameters of 256, 128, and 64 roughly correspond to angular resolutions of approximately\footnote{Since the shape of the HEALPix pixels are not all the same, this number is approximated.} 13.74, 27.48, and 54.96 arcmin.

First, we note that at 132.63 MHz, the fringe spacing is $\approx34.5$ arcmin for the 226-m EW baseline, the longest baseline in our simulation, whereas the approximate resolution of the HEALPix grid at NSIDE=256 is 13.74 arcmin. This means that the fringe patterns are sampled at a slightly higher rate than the Nyquist rate, and thus the discretised diffuse model should capture most of the information in the original GSM map, which should translate into the most accurate delay spectrum for the baselines that we simulate. We have run a test with a diffuse model of higher NSIDE and found it to be in numerical agreement with the NSIDE=256 case.

As we lower the resolution of the diffuse model, the discretised source density reduces, as well as the sampling of the fringes, resulting in aliasing and leading to errors in the power spectrum. This is apparent in the case of NSIDE 64 and 128 for 226-m EW and 143-m NS baselines. The latter corresponds to a fringe spacing of $\approx54.6$ arcmin, and thus provides a close to Nyquist sampling, but, as we mentioned, the HEALPix pixel resolution is not exact, and this leads to small errors in the delay spectrum as shown.

For the 14-m baseline, the fringe spacing is $\approx5^\circ$, which should be coarse enough to be well sampled by all HEALPix resolutions in this test, and the resulting delay spectrum matches visually at all NSIDEs. 
However, their fractional residuals are still $\sim10^{-2}$ or $~1\%$ at all delays. This may or may not be accurate enough depending on the use case. 
In general, a good rule of thumb is to sample the sky model beyond the Nyquist rate.

\subsection{Imaging \matvis{} Visibilities}

Although \matvis{} was primarily motivated by the need to validate 21\,cm drift-scan-style delay-spectrum experiments, and currently does not support simulating tracking observations, imaging the output visibilities is certainly possible. 
In this case, the main concern is how errors from our fast coordinate rotation scheme manifest in the resulting images.

\begin{figure*}
    \centering
    \includegraphics[width=\textwidth]{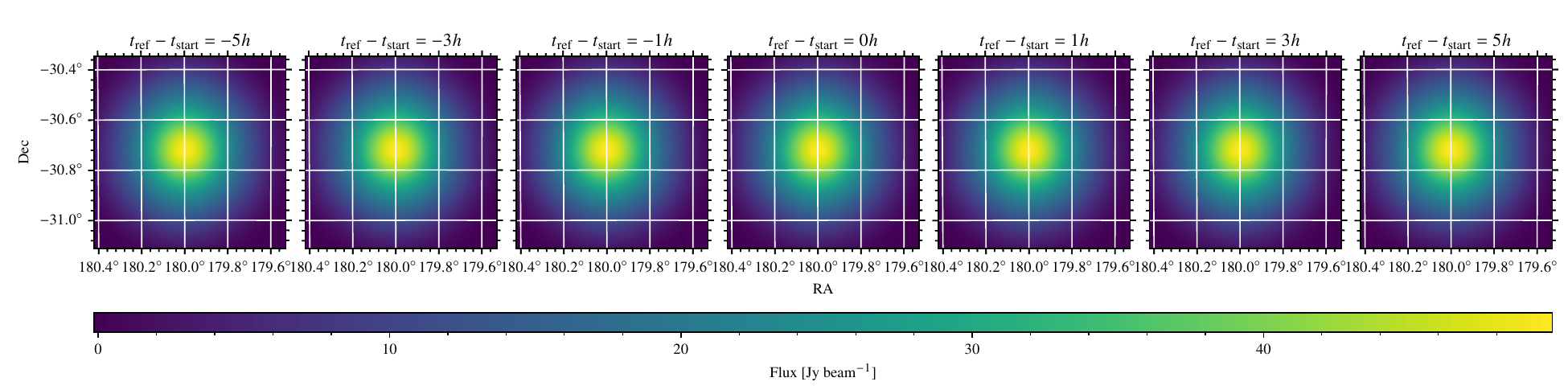}
    \includegraphics[width=\textwidth]{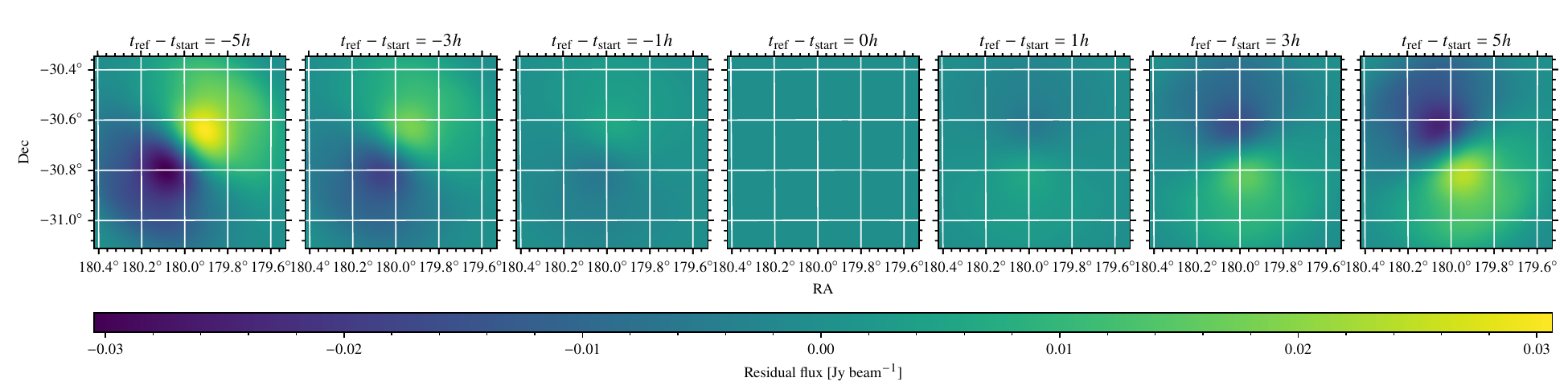}
    \caption{Demonstration of the effects of errors in \matvis{} `fast' coordinate transformation on the dirty images of its output visibilities. Here, a snapshot of a single 100-Jy point source transiting near the zenith of the full HERA array was simulated, and then converted to a dirty image with the \texttt{tclean} task on \texttt{CASA}. The top row shows the imaging results, for which each column is derived from a simulation with a different \matvis{} coordinate transformation reference time relative to the simulated time. The bottom row shows the residual differences of each column to the middle column, for which the start and reference times are the same, thus providing the most accurate source position. As expected, the source position error due to \matvis{} coordinate transformation scheme leads to a position shift of the point source in the image.}
    \label{fig:img_test}
\end{figure*}

We simulate multiple 5-second snapshots of a single, 100-Jy, point source at RA$=12$ hour and Dec$=33.6^{\circ}$ transiting near the zenith of the full, 350-antenna, HERA array (maximum baseline $\approx800$ m).
In each case, the snapshot time remains the same, but we modify the coordinate transform `reference time'. We form dirty images with the \texttt{tclean} task in \texttt{CASA} with 5'' pixel size. The top row of Figure~\ref{fig:img_test} shows the resulting images, which are visually indistinguishable. 
The central column of the figure corresponds to the simulation in which the simulation time matches the reference time.

However, the bottom row shows each image as a residual to the central column -- i.e. the difference between assuming an offset reference time compared to the a precise coordinate transformation.
Here, it is apparent that the point source is slightly shifted as the reference time diverges from the simulation time.
Similar to how this affects the delay spectrum in Figure~\ref{fig:dps_ref_time}, the further the difference between the observation time and the reference time, the larger the error. 

This is expected as source position errors will lead to phase errors and will thus result in shifting or smearing of the source in the image. This error will likely be more acute when simulating an array with longer baselines. We leave this demonstration as a precaution for the readers, and obviously, the best practice is to simulate one snapshot at a time and set the reference time to the observation time.

%============================================================================================
\section{Computational Performance and Optimisation} 
\label{sec:scaling}
%============================================================================================

\begin{figure*}
    \includegraphics[width=\textwidth]{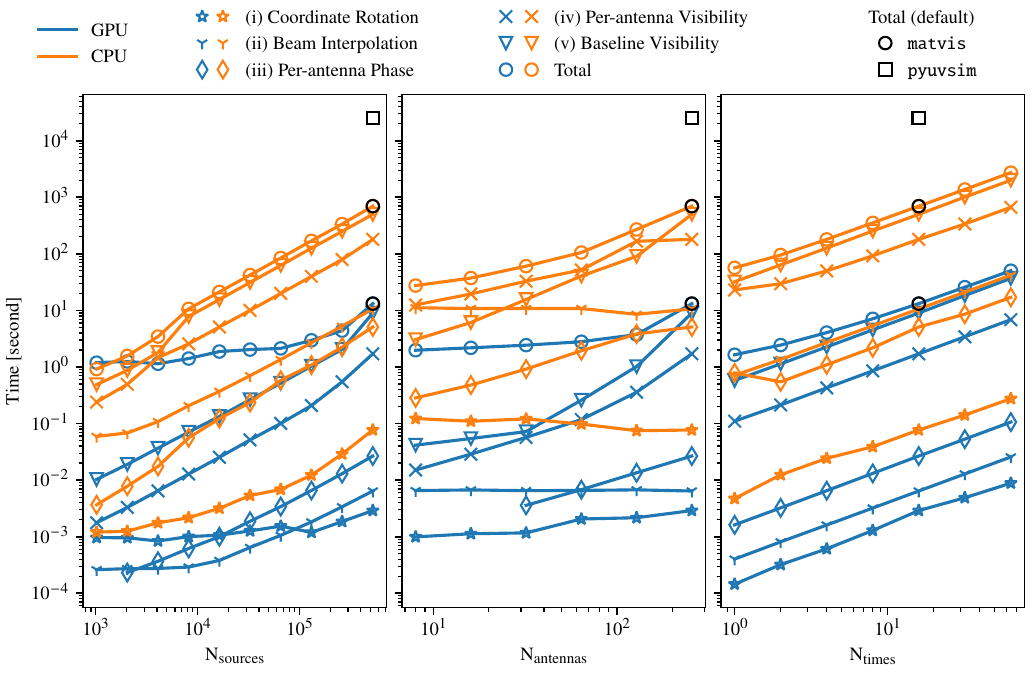}
    \caption{Scaling test results of the CPU and GPU implementation of \matvis{}. Orange and blue colours indicate CPU and GPU results, while the different markers indicate steps in the calculation as described in Section~\ref{sec:algorithm}. The wall time for the default simulation parameters   ($N_{\mathrm{antennas}}=256,\ N_{\mathrm{sources}}=2^{19},\ N_{\mathrm{times}}=16$) are shown in a black circle markers for \matvis{} and a black square  marker for \pyuvsim{} for comparison. Note that scaling tests of \pyuvsim{} indicate that it scales similarly to \matvis{} except at very small simulation sizes where it has some higher overheads. }
   \label{fig:scaling_test}
\end{figure*}

An important feature of \matvis{} is its computational performance. 
Indeed, the consideration of performance is the driving motivation behind the matrix-based algorithm presented in Section~\ref{sec:algorithm}, and the structuring of the Python implementation in a way that GPUs can be used as a drop-in replacement for CPUs.
In this section, we will look at the relative contribution to overall run-time from each of the algorithm steps, determining the bottleneck(s). We will also explore how this run-time scales with different simulation parameters, numbers of cores, and how much acceleration is afforded by the GPU. We will also consider the memory footprint of the algorithm and associated concerns, and finally we will take a deeper look at the largest bottleneck---the matrix-matrix multiplication of step (v)---considering its theoretical complexity and potential optimisations.

\subsection{Runtime Performance}

Figure~\ref{fig:scaling_test} summarises the runtime performance of \matvis{}.
Specifically, it shows the timing of a number of simulations with different numbers of source $N_{\rm sources}$, antennas $N_{\rm antennas}$, and times/snapshots $N_{\rm times}$, using the script \verb|matvis profile| included in the package. 
The script generates simulation configurations that take advantage of the fact that the performance cannot depend on the properties or locations of sources, the locations of antennas, or the precise frequencies or times of observation: only their total number.
In addition to these dimensions, the performance will depend on the number of unique beams, $N_{\rm beams}$, and whether they are analytic or simulated (and if simulated, the size of the grid onto which they are simulated), as well as the number of feeds per antenna, $N_{\rm feed}$ and whether the beams are $E$-field beams or not.

In Figure~\ref{fig:scaling_test} we consider only the case where $N_{\rm beams} = 1$, i.e. all antennas have the same beam (which is a common assumption). This beam is a complex-valued $E$-field beam evaluated on a 1-degree regular mesh for two feeds. 
To be representative of common simulation setups, the sources are distributed uniformly across the sky, which means that at any particular integration, half of them are below the horizon. 
In the current implementation of \matvis{}, these sources are ignored (for that given integration), which potentially impacts performance.

Figure~\ref{fig:scaling_test} shows the impact of three parameters.
Our `default' value for each is $N_{\rm sources}=2^{19}\approx 525$k, $N_{\rm antennas}=256$ and $N_{\rm times} = 16$.
Keeping two parameters fixed at their default, we run the third parameter over a logarithmic grid: $N_{\rm sources} \in 2^{\{10, \dots, 19\}}$, $N_{\rm antennas} \in 2^{\{3,\dots,8\}}$ and $N_{\rm times} \in 2^{\{0,\dots,7\}}$.
Run-times are shown for the CPU (orange) and GPU (blue), with the default run shown in black. 
We also show the run-time for the default set using \pyuvsim{} as a black square.
Different marker symbols indicate steps of the algorithm (cf. Section~\ref{sec:algorithm}), which we will discuss in more detail momentarily. 
These profiling simulations were performed on the Ilifu cloud-computing cluster\footnote{\url{https://www.ilifu.ac.za/}} in Cape Town, South Africa. The CPU profiles are computed on a 2.6GHz Xeon Processor using a single core for each run. The GPU profiles are computed on NVIDIA V100 GPU.
Note that the results shown are for a single frequency; there is almost perfect parallelism between frequencies, and virtually no overhead from computing a single frequency in a single job. There are two caveats to this: firstly, the computation of the array $\boldsymbol{\mathsf{R}}$ for all time-steps can be done once for all frequencies, and secondly, writing out the simulated data to disk can be more efficient when frequencies are grouped together.
The computation of $\boldsymbol{\mathsf{R}}$ is included in the total time shown in Figure~\ref{fig:scaling_test}, contributing to the small overhead. I/O overheads from reading the beam are included, but not from writing the results to disk. 
The total run-time (for both CPU and GPU) is calculated by simply measuring the wall-time of the entire calculation. 
The run-time of individual algorithm steps is measured on the CPU by using the \texttt{line\_profiler} Python package. 
On the GPU, since each step corresponds to a particular CUDA kernel, we use the \texttt{nvprof} utility which specifically measures performance of kernels at runtime. We note that the sum of the run-times of individual steps should be strictly \textit{less} than the total time, which also takes into account overheads outside the main loop of the algorithm as well as internal overheads (e.g. Python for-loop overhead).

The first thing we notice is that the scaling with $N_{\rm times}$ is almost perfectly linear for all steps of the calculation. This is exactly as we expect, since each step is run once per loop and the overhead outside the loop is very minimal compared to the run-time of the default parameters.
The scaling with $N_{\rm antennas}$ and $N_{\rm sources}$ is far more interesting, and varies amongst the different algorithm steps.

A second obvious takeaway is that the GPU accelerates the computation by a factor of $\sim100$ for the larger simulations tested here. This is clearly dependent on the properties of both the CPU and GPU used, but should be a good rule-of-thumb.
Even the slower CPU implementation is $\sim40\times$ faster than the \pyuvsim{} simulator in its current form.

For both GPU and CPU, the major component in terms of run-time is the matrix-matrix multiplication, step (v). For the majority of problem sizes on the GPU, this step is a factor of $\sim 5$ more expensive than the next-most-expensive step, which is generally the calculation of the per-antenna matrix $\boldsymbol{\mathsf{Z}}$, i.e. step (iv). 
Interestingly, on the CPU this is only true for the largest $N_{\rm antennas}$ considered here: for smaller $N_{\rm antennas}$, step (iv) is dominant. The scaling of these two steps with $N_{\rm antennas}$ is expected to be different, with step (v) naively scaling more strongly than step (iv), which explains this cross-over for large $N_{\rm antennas}$.
Another caveat to this picture is that on the GPU, when $N_{\rm sources} \lesssim 10^5$ \textit{or} $N_{\rm antennas} \lesssim 128$, the dominant contribution to run-time is in fact \textit{overheads}, which for this setup appear to be about one second. While some of this overhead is from outside the time-loop, inspection of the profile reveals that much of the overhead is from within the time-loop (between GPU operations), and therefore will scale with $N_{\rm times}$. This will present an issue for simulations with small $N_{\rm ant}$ with large $N_{\rm times}$, but can perhaps be overcome with further optimisations to the Python code.
A final caveat is that while the beam interpolation is a negligible component here, this is partially because we have used $N_{\rm beam}=1$, whereas in principle it could scale with $N_{\rm antennas}$ if all antennas are simulated with varying beams. If this is the case, it would become the \textit{dominant} step on the CPU up to very large $N_{\rm antennas}$ (where step (v) would overtake due to its hyperlinear scaling), and on the GPU it would be similar to step (iv) in its impact.

Since the algorithm's performance is indifferent to whether baseline vectors are unique or not, the scalings shown in Fig. \ref{fig:scaling_test} hold both for highly-redundant (e.g. HERA) and less redundant arrays (e.g. MWA, SKA). 
The performance of \matvis{} for next-generation SKA-LOW with 512 antennas can thus be predicted to be approximately 3 seconds per channel per time-stamp for a sky model with $\sim 5\times10^5$ sources.

\subsection{Multi-Core CPU Performance}
\begin{figure}
    \centering
    \includegraphics{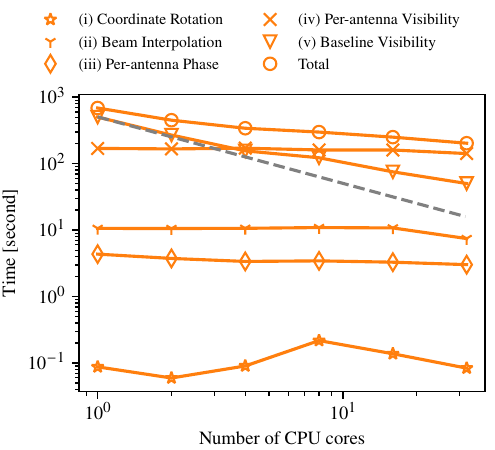}
    \caption{Scaling test results for multi-cores CPU performance of \matvis{}. The grey dashed line indicates a linear scaling.}
    \label{fig:multicores}
\end{figure}

Modern linear algebra and interpolation routines, such as those in \texttt{NumPy} and \texttt{SciPy}, can take advantage of multiple CPU cores through multi-threading. To measure this performance, we run the profiling script with the default parameter set giving 2, 4, 8, 16 and 32 number of CPUs per job. We compile \texttt{NumPy} with the Intel Math Kernel Library, and set the \texttt{MKL\_NUM\_THREADS} environment variable to the number of CPUs used in each job to specify the number of threads. Figure~\ref{fig:multicores} shows the scaling test results. Different markers correspond to different steps in \matvis{} as in Figure~\ref{fig:scaling_test}. The grey dashed line indicate linear scaling (i.e. -1 slop). Notice that only step (v), which forms the per-baseline visibilities through the matrix product of the per-antenna visibilities, benefits from multi-threading. The scaling is linear until $\approx 4$ CPUs, and then it flatten off. While further CPU cores do continue tp speed up step (v), step (iv) becomes the bottleneck by that point since it is simply performing complex array arithmetic, and it does not benefit from multi-threading. Thus, we recommend that each job is given 2-4 cores when simulating with the CPU implementation. 

\subsection{Optimisation}
\label{sec:detailed-optimization}
Given the above observations, the crucial components to optimise for the majority of current use cases are steps (iv) and (v) -- and particularly (v) as the number of antennas grow larger in modern experiments (for example, SKA Phase II will have 512 antennas, double the maximum tested in this section). 

Step (iv) naively has complexity $\mathcal{O}(N_{\rm ant}N_{\rm src})$, and this is borne out in Figure~\ref{fig:scaling_test} and \ref{fig:multicores}. It requires calculation of a complex-exponential for each of the terms, which can be quite computationally expensive. However, it is also naively parallelism over either the source or antenna axes. Which is more convenient depends on the memory layout of the input and output arrays.

Step (v) is a more interesting problem.
Naively, its complexity is $\mathcal{O}(N_{\rm ant}^2 N_{\rm src})$. 
This scaling applies inescapably to traditional simulators that compute the RIME as a set of $N_{\rm baseline}$ tasks, of each which computes Equation~\ref{eq:matvisrime}, since $N_{\rm baseline} \propto N_{\rm ant}^2$.
However, our matrix-based approach makes use of highly-optimised algorithms for one of the most common operations in all computing: the matrix-matrix product.
The matrix $\boldsymbol{\mathsf{Z}}$ has shape $N\times M$, where $N = N_{\rm ant}N_{\rm feed} \sim 10^{2-3}$ and $M = N_{\rm ax}N_{\rm src} \sim 10^{5-6}$.
For \textit{square} matrix multiplication ($M \equiv N$), the optimal complexity is known to be $\mathcal{O}(N^{\omega})$, with $\omega \approx 2.37$ or better~\citep{williams.etal.2023}. For more generic rectangular matrix multiplication, \citet{legall.urrutia.2017} places $\omega(k) < 2 + k$ as an upper bound for multiplying an $N \times N^k$ matrix by an $N^k \times N$ matrix for $k \leq 5$ (e.g.  $\omega \leq 6.16$ for $k = 5$).  Thus, an optimised matrix multiplication algorithm should perform better than $\mathcal{O}(N_{\rm ant}^2)$ even in the realistic situation where $N_{\rm src} \gg N_{\rm ant}$. 
We have checked several commonly-used Python routines, and found the complexities in the range of $\mathcal{O}(N_{\rm ant}^{1.3-2})$, reaffirming this claim.  
Thus, the algorithm presented here has every chance of being unbeatable for simulating current and future large-$N_{\rm ant}$ experiments. 

In detail, some choices need to be made about the implementation of the matrix-matrix product. 
For both the CPU and GPU the current implementation uses the \texttt{ZGEMM} routine from \texttt{BLAS} to perform the multiplication. 
Specifically, on the CPU, the call is made via the \texttt{numpy.dot} function, which internally uses a backend implementation of \texttt{BLAS} that is configurable by the user. In Figure~\ref{fig:scaling_test} and \ref{fig:multicores}, we used the MKL backend. On the GPU, we use the \texttt{CuBLAS} implementation. Comparison to other implementations (e.g. \texttt{ArrayFire}\footnote{\url{https://arrayfire.org}} and \texttt{Magma}\footnote{\url{https://icl.utk.edu/magma/}}) is ongoing. 

Besides the choice of implementation package, there are further optimisations worth considering. 
For example, the product $\boldsymbol{\mathsf{V}} = \boldsymbol{\mathsf{Z}}\boldsymbol{\mathsf{Z}}^\dagger$ contains hermitian symmetry that in principle can be exploited by the \texttt{ZHERK} routine in \texttt{BLAS}, reducing the computations by a factor of two. 
However, we have found this approach to be less performant than the full \texttt{ZGEMM} for a range of realistic problem sizes.
This is most likely due to the highly-irregular size of the matrices involved (i.e. $M \gg N$).

Another set of symmetries that this approach does not exploit are \textit{baseline redundancies}: two baselines: $ij$ and $kl$ have equal visibilities if their beams are the same ($A_i \equiv A_k$ and $A_j \equiv A_l$) and their baseline vector is the same: $\boldsymbol{b}_{ij} = \boldsymbol{b}_{kl}$, regardless of the absolute positions of the antennas. 
For compact arrays like HERA---under the common assumption of ideal beams---the fraction of baselines which are redundant is very large ($\gtrsim 95\%$), which means that $\texttt{ZGEMM}$ is naively performing a factor of $>20-40$ more calculations than necessary. 
We consider this problem in a little more detail in Appendix \ref{app:enhancement}, noting here only that for very large $N_{\rm ants}$, the optimal scaling of the matrix-multiplication will still win, even with the redundant calculations.
In particular, \matvis{} should be highly performant for upcoming large-$N$ non-redundant arrays such as the SKA.

\subsection{Memory Considerations}
\label{sec:memory}
The way that the \matvis{} algorithm is structured means that all steps inside the algorithm loop need only be stored in memory for a single snapshot, and only the final result needs to be kept for all snapshots. This final output is a complex array of size $N_{\rm ant}^2 N_{\rm feed}^2 N_{\rm time}$. For the common case of $N_{\rm feed}=2$ and double-precision calculations, this is an array of size
\begin{equation}
    {\rm size}(\boldsymbol{\mathsf{V}}) \approx \left(\frac{N_{\rm ant}}{128}\right)^2 \frac{N_{\rm time}}{1000}\ \ {\rm GB}.    
\end{equation}
Inside the loop, the largest array is $\boldsymbol{\mathsf{Z}}$, with 
\begin{equation}
    {\rm size}(\boldsymbol{\mathsf{Z}}) \approx \frac{N_{\rm ant}}{128}\frac{N_{\rm src}}{130,000} \ \ {\rm GB}.
\end{equation}

For the GPU implementation, this latter size is very important, as it is the dominant source of device-based memory. 
If $N_{\rm src}$ is very large, the peak memory (and device memory) can be easily reduced by splitting the sources into chunks, for minimal overhead (since the matrix-matrix calculation scales linearly with $N_{\rm src}$), and \matvis{} does this automatically.

Likewise, the total peak memory can be controlled by chunking the number of times, which incurs minimal overhead (according to Figure~\ref{fig:scaling_test}) as long as $N_{\rm ant}$ and $N_{\rm src}$ are reasonably large. 

%============================================================================================
\section{Conclusions} \label{sec:conclusions}
%============================================================================================

In this paper, we introduced the \matvis{} visibility simulator and its matrix-based algorithm for calculating the RIME. By first calculating per-antenna responses and then expanding them to visibilities through a simple outer product, \matvis{} is able to utilise highly-efficient matrix operations, especially those implemented for GPUs, to accelerate the calculation of the RIME. The algorithms assume a point-source approximation for the sky model, treating each of the components in the model as an unresolved point source with a defined total flux density at a given frequency. 
To simulate diffuse sky emission or extended sources, they must first be discretised into an ensemble of point sources.
To improve the performance of \matvis{}, equatorial-to-horizontal coordinate transformations are performed as simple rotations, which introduces a slight degradation of accuracy if simulating a series of snapshots over a long interval.
To mitigate this degradation, we introduce a computational trick that uses accurate \texttt{astropy}-based transformations to partially correct the errors in the fast geometric rotations, providing a reference time for the correction. This `fast` coordinate transformation scheme yields the same results as astropy, and thus the most accurate source positions, when simulating from a single time snapshot\footnote{We note again that this approximate scheme is no longer required in versions of \matvis{} greater than 1.3.0, and fast-and-accurate schemes can be used instead.}. 

The primary use case of \matvis{}, which is also the motivation for its development, is to generate mock visibilities from many-element 21~cm arrays, particularly HERA, and for validating the power spectrum analysis pipelines of these experiments. 
However, the presented algorithm and the \matvis{} Python package itself are general and can be applied to other use cases that require mock visibilities. 
We have also developed the \texttt{hera\_sim} package that provides HERA-specific systematic models and a unified interface for visibility simulators with already built-in wrappers for \matvis{} and \pyuvsim{}, a reference simulator that prioritises accuracy and has been tested against analytical calculations. 
The \texttt{hera\_sim} wrapper is our recommended way to get started using \matvis{}.

We validated \matvis{} for its 21~cm analysis use cases, particularly by looking at the dynamic range of the delay spectrum and power spectrum estimates from the simulations. We also compared \matvis{} against \pyuvsim{} to test its accuracy when simulating single time snapshots.
For the delay spectrum, we simulated a single time snapshot from a mock full-sky point source catalogue derived from the source count distribution of the GLEAM \citep{franzen.etal.2019} catalogue, and a GSM diffuse sky model \citep{gsm.2008} with both simulators, and show that delay spectrum derived from these foreground simulations are well-matched to \pyuvsim{} to $\sim10^{-10}$ in fractional residuals. 
However, when simulating long-duration observations for the purposes of calibration, we recommend that the simulation be split into $\sim15$\,minute time chunks to alleviate the errors from the approximate coordinate transforms. 
We also show that the obtained foreground delay spectrum maintains a peak-to-floor dynamic range $\gtrsim10^{12}$ regardless of the reference time used for \matvis{}' `fast' coordinate transforms. 
We also show that simulated visibilities drawn from a 21\,cm-only sky model, with known input power spectrum, produce power spectra that closely match the input power, within sample variance.
These tests support \matvis{}  as a suitable simulator for the validation of 21~cm delay-spectrum pipelines. 
Furthermore, since residuals against \pyuvsim{} are $\lesssim 10^{-10}$ when simulating a single time snapshot, \matvis{} simulations performed in this way (one snapshot at a time) are of sufficient accuracy for the purposes of calibration of wide-field 21~cm data (given a suitable sky model). 

We also demonstrate the fundamental limitation of \matvis{}'s point-source approximation (and other simulators that make this assumption) when simulating from discretised diffuse emission at long baselines. 
In short, the resulting delay spectrum (or any measurement) derived from the simulated visibility will contain errors due to aliasing in the fringe term of the RIME. The rule of thumb is to make sure that the diffuse sky is discretised on a coordinate grid, e.g HEALPix, at least twice the angular resolution of the longest baseline in the simulation, i.e. following the Nyquist–Shannon sampling theorem.

There are currently a few limitations to \matvis{}: (i) it can only simulate drift-scan observations; (ii) it can only simulate unpolarised (Stokes I) sky models, although it account for polarised beam effects; and (iii) it cannot simulate just a subset of the pairs of a given set of antennas, for example just the non-redundant baselines in HERA. 
One way to enable `tracking observations' to address (i) is to perform a post-processing step that rephases the output visibilities to a chosen phase centre.
As for (ii), the algorithm presented in Equation~\ref{eq:fullz} and \ref{eq:zz}, which is a general case of the current \matvis{} algorithm, is already capable of handling a fully polarised sky.
However, implementing it would require an additional tensor summation step, which must be tested and will affect the performance.
Solving (iii) would require changing (or adding) how the visibility is computed in Equation~\ref{eq:matvisrime} to support a partial antenna matrix.
Lastly, although not an essential limitation, the fast coordinate transformation approximation is a rough edge that can currently only be overcome by a slow outer-loop chunking simulations over snapshot times. This limitation is currently being removed so that arbitrarily high-precision long-duration simulations will be possible with future versions of \matvis{}. 
We leave these as features for future versions of \matvis{} and welcome contributions from the community.

%--------------------------------------------------------------------------------------------
\section*{Acknowledgements}
%--------------------------------------------------------------------------------------------
This material is based upon work supported by the National Science Foundation under grants \#1636646 and \#1836019 and institutional support from the HERA collaboration partners. 
This research is funded in part by the Gordon and Betty Moore Foundation through Grant GBMF5212 to the Massachusetts Institute of Technology.
HERA is hosted by the South African Radio Astronomy Observatory, which is a facility of the National Research Foundation, an agency of the Department of Science and Innovation. 

P.~Kittiwisit acknowledges the financial assistance of the South African Radio Astronomy Observatory (SARAO; \url{www.sarao.ac.za}). This result is part of a project that has received funding from the European Research Council (ERC) under the European Union's Horizon 2020 research and innovation programme (Grant agreement No. 948764; P.~Bull). P.~Bull and H.~Garsden acknowledge support from STFC Grants ST/T000341/1 and ST/X002624/1. S. G. Murray has received funding from the European Union’s Horizon 2020 research and innovation programme under the Marie Skłodowska-Curie grant agreement No. 101067043.  C. Cain is supported by the Beus Center for Cosmic Foundations.  This work used Bridges-2 at the Pittsburgh Supercomputing Centre (PSC) through allocation PHY210042 from the Advanced Cyberinfrastructure Coordination Ecosystem: Services \& Support (ACCESS) program, which is supported by National Science Foundation grants \#2138259, \#2138286, \#2138307, \#2137603, and \#2138296. We acknowledge the use of the ilifu cloud computing facility -- \url{www.ilifu.ac.za}, a partnership between the University of Cape Town, the University of the Western Cape, Stellenbosch University, Sol Plaatje University, the Cape Peninsula University of Technology and the South African Radio Astronomy Observatory. The ilifu facility is supported by contributions from the Inter-University Institute for Data Intensive Astronomy (IDIA - a partnership between the University of Cape Town, the University of Pretoria and the University of the Western Cape), the Computational Biology division at UCT and the Data Intensive Research Initiative of South Africa (DIRISA).

%%%%%%%%%%%%%%%%%%%%%%%%%%%%%%%%%%%%%%%%%%%%%%%%%%
\section*{Data Availability}

\matvis{} and \texttt{hera\_sim} have been developed or contributed by members of the HERA collaboration. \pyuvsim{} and \texttt{pyuvdata} are developed and maintained by the Radio Astronomy Software Group (RASG). All packages mentioned are publicly available on GitHub (see Section~\ref{sec:package} for details). Visibility files from Section~\ref{sec:validation} can be made available upon request to the corresponding author.

%%%%%%%%%%%%%%%%%%%% REFERENCES %%%%%%%%%%%%%%%%%%

\bibliographystyle{rasti}
\bibliography{matvis} 

%%%%%%%%%%%%%%%%% APPENDICES %%%%%%%%%%%%%%%%%%%%%

\appendix

\section{Utilising Baseline Redundancies}
\label{app:enhancement}

Let the matrix $\bs{Z}$ have shape $N\times M$, where $N = N_{\rm feed}N_{\rm ant}$ and $M = N_{\rm ax}N_{\rm src}$.
Then the matrix $\bs{V} = \bs{Z}\bs{Z}^\dagger$ has shape $N\times N$.
As discussed in Section~\ref{sec:detailed-optimization}, one of the disadvantages of the current \matvis{} implementation is that all $N^2$ antenna-feed pairs must be simulated, including those that are `redundant'. 

To be clear, this redundancy occurs when two baselines, $ij$ and $kl$, have equivalent beams and baseline vectors ($\boldsymbol{b}_{ij} = \boldsymbol{b}_{kl}$).
We can split the baselines of any array into a number $N'_u$ of sets of baselines, where each set contains all $n_i$ baselines that are redundant with each other, such that $\sum_i n_i = N_{\rm ant}(N_{\rm ant}+1)/2$ (the sets don't contain reverse-pairs since their visibilities are simply the conjugate of the reverse).
Letting $N_u = N_{\rm feed}^2 N'_u$ be the number of non-redundant elements of $\boldsymbol{\mathsf{V}}$, we can write the ``non-redundancy fraction'',
\begin{equation}
    \xi = N_u / N^2,
\end{equation}
which is very low for highly-redundant compact arrays like HERA. Thus, calculating the full $N^2$ matrix is performing far more calculations than strictly necessary.

\begin{figure}
    \centering
    \includegraphics[width=\columnwidth]{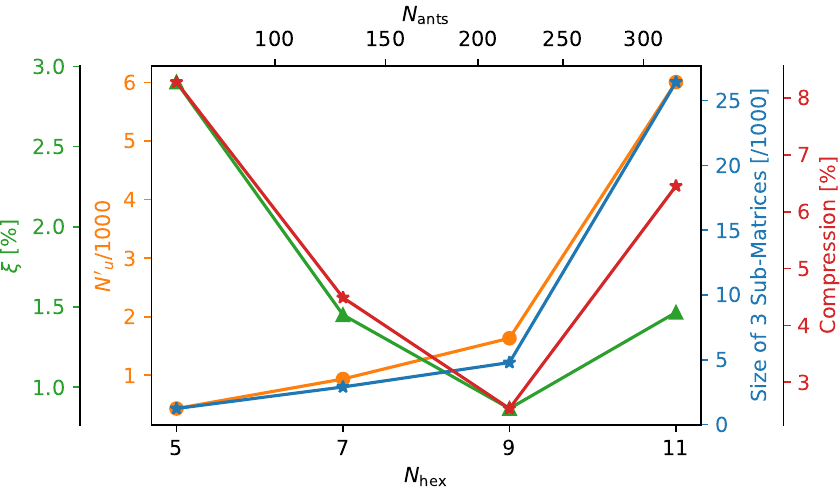}
    \caption{Characteristics of the `HERA-like' arrays used in this appendix. The $x$-axis denotes the size of the array, defined by $N_{\rm hex}$ (the number of antennas along a side of the hexagon) and $N_{\rm ants}$. 
    The orange and green curves (left spines) show characteristics of the arrays: the number of unique pairs (orange) and the non-redundancy fraction, $\xi$. 
    The blue and red curves (right spines) show indicate properties of the sub-matrix algorithm outlined in this appendix: the total size of the sub-matrices (in this case, 3 sub-matrices were used) in blue, and this size as a fraction of the full matrix.
    Unique pairs grow almost linearly with $N_{\rm ants}$.
    }
    \label{fig:hera_nants}
\end{figure}

In this appendix, we consider a few alternative approaches aimed at improving performance in this situation. 
Far from being a fully-fledged examination, we will limit our focus to a few ideas, and test their performance only on `HERA-like' arrays, by which we mean that the antennas are arranged in a hexagonal core, split into three parallelogram shards (reducing redundancy) with $N_{\rm hex}$ antennas along a side. In this setup, $\xi$ varies between $\sim$1-3\%, and 
$N'_u$ varies between 500-6000 (see Figure \ref{fig:hera_nants}).

\begin{figure*}
    \centering
    \includegraphics[width=\textwidth]{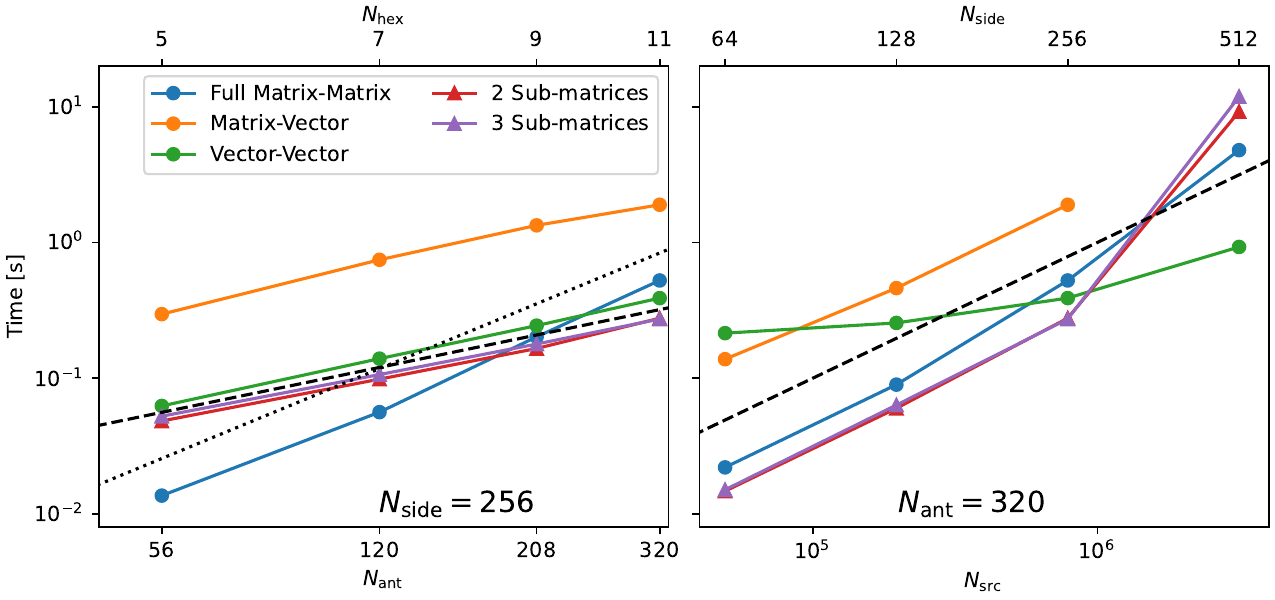}
    \caption{Performance characteristics of four different approaches to calculating $\bs{V}$. The LHS panel shows performance as a function of $N_{\rm ant}$, where the array is a `HERA-like' array with characteristics shown in Figure \ref{fig:hera_nants}. Here $N_{\rm side}=256$ is used. On the RHS, $N_{\rm ant}$ is fixed at 320 and the performance as a function of $N_{\rm src}$ is shown. The full matrix-matrix method adopted by \matvis{} is in blue, the matrix-vector approach in orange, the vector-vector dot product approach in green, and the more general sub-matrix approach in red and purple (for different numbers of sub-matrices). Black dashed line indicates linear scaling, and black dotted indicates quadratic scaling.}
    \label{fig:perf_test}
\end{figure*}

Figure \ref{fig:perf_test} shows in blue the performance of the default implementation of \matvis{}: the full matrix-matrix (MM) calculation. On the LHS the performance as a function of $N_{\rm ant}$ is shown, and on the RHS as a function of $N_{\rm src}$. The black-dashed line is an indication of linear scaling with each parameter, whereas the black-dotted line indicates scaling with $N_{\rm ant}^2$. It is clear that the GPU implementation used here scales close to $N_{\rm ant}^2$.

The most straightforward way to use the redundancy of the array is to compute only the $N_u$ vector-vector (VV) dot products required within a for-loop.
This has complexity $\mathcal{O}(N_u M \equiv \xi N^2 M)$, which is better than the $\mathcal{O}(N^2M)$ that the full matrix-matrix calculation exhibits in Figure \ref{fig:perf_test}.
%Assuming $\xi$ is independent of $N$, this is clearly worse scaling to high-$N$ than can be achieved with the full matrix-multiply (recall that this is $\mathcal{O}(N^{1.3-2}M)$, cf. Section~\ref{sec:detailed-optimization}).
On the other hand, we expect this to have some overhead associated with the for-loop, and this is borne out in Figure \ref{fig:perf_test}, where this method is shown in green.
For small $N_{\rm ant}$, the overhead associated with the for-loop is large enough that this method is $\sim 6\times$ slower than the MM method. However, it grows with $N_u$ instead of $N^2$, which in the HERA-like case considered here is close to $N_u \propto N$. For the full HERA array, the VV loop is more performant than the MM calculation.
Considering the performance against $N_{\rm src}$, the VV loop has very poor performance for small $N_{\rm src}$, and only starts to hit the theoretical linear growth for $N_{\rm src} \gg 10^6$. This must be due to low utilisation of the many GPU cores: each VV dot product is performing $N_{\rm src}$ calculations split between thousands of cores. 

A more general alternative approach is to deconstruct $\bs{V}$ into a set of smaller matrices $\{\bs{V}'_{x}\}$, for which each $\bs{V}'_x$ contains some non-zero number of unique pairs in $\bs{V}$, and is a smaller yet large enough matrix to take advantage of modern matrix multiplication routines,
\begin{equation}
    \label{eq:Vpx}
    \bs{V}'_{x} = \bs{A}_x \bs{B}^\dagger_x.
\end{equation}
Here, $\bs{A}_x$ and $\bs{B}_x$ are both formed from (possibly non-adjacent) full rows of $\bs{Z}$, having shape $N_A \times M$ and $N_B \times M$ respectively.
The rows chosen for $\bs{B}_x$ correspond to the columns of $\bs{V}'_x$.
We can then attempt to optimise the sets $\{\bs{A}_x\}$ and $\{\bs{B}_x\}$ to yield the best possible computation time for $\{\bs{V}'_x\}$ (and thus for $\bs{V}$) given a particular set of unique antenna pairs. This optimisation is obviously dependent on the antenna distribution. 
Note that both the vector dot-product approach described above and the full matrix product $\bs{V}=\bs{Z}\bs{Z}^\dagger$ are special cases of Equation~\ref{eq:Vpx}, where the number of matrices in $\{\bs{V}'_x\}$ is $N_u$ and $1$, respectively.  

For the full HERA array, the number of possible choices for $\{\bs{V}'_{x}\}$ is extremely large (since $N = 700$)\footnote{There are $N_u$ choices of \textit{number} of matrices in the set, and for each such choice there a multitude of ways to configure the shape of each sub-matrix.}, which makes Equation~\ref{eq:Vpx} a daunting optimisation problem.  
However, the structure of the matrix can potentially lead to some useful ideas.

One such intermediate idea is for each $\bs{A}_x$ to be a $1\times M$ matrix, and each $\bs{B}_x$ to include only the antennas for which the single antenna in $\bs{A}_x$ appears in non-redundant pairs. This computes \textit{only} the redundant pairs, like the VV method above. However, it is liable to utilise vectorisation better than the VV approach as it includes fewer but larger matrix-vector (MV) products.
On the other hand, memory management may be a problematic overhead in this approach since for each of the $N$ MV products, the matrix needs to be `formed' by coalescing memory from non-adjacent rows in the full $\bs{Z}$ matrix. 
This is borne out in Figure \ref{fig:perf_test}, where this method is shown in orange. While the scaling of the MV method with both $N_{\rm ant}$ and $N_{\rm src}$ is what we expect (and is therefore better than MM for some large $N$ where $\xi$ is very small), it is much slower than all the other methods, due to both the overhead in the for-loop ($N_{\rm ant}$ iterations are required) and in the movement of data on the GPU to form $\bs{B}_x$.

\begin{figure*}
    \centering
    \includegraphics[width=\textwidth]{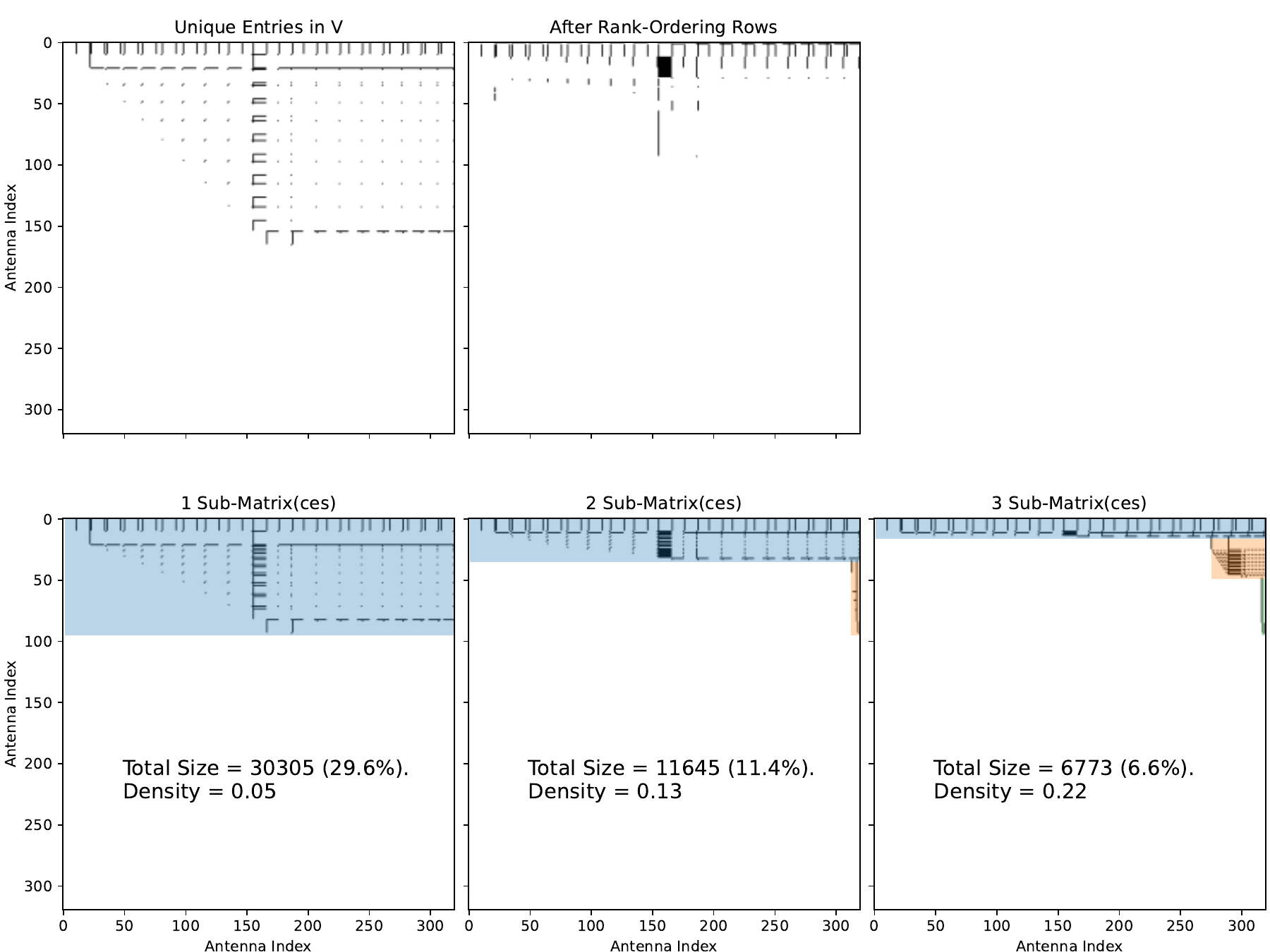}
    \caption{A visualisation of our sub-matrix procedure for taking advantage of antenna-pair redundancies in HERA.  
    The plot is meant to be read from top-left to bottom-right.
    The top-left represents $\bs{V}$ for HERA-320, where black pixels indicate unique baselines. These black pixels are less than 3\% of the pixels. In the top-right, the rows have been re-ordered such that the top row has the most black dots, and the bottom row the least. 
    On the bottom panel, we show the result of our algorithm for partitioning the rows into different numbers of subsets (1, 2 and 3 from left to right). In each case, after deciding on a vertical extent for the matrix (coloured region), the columns \textit{within} the matrix are shuffled such that columns occupied by any row within the matrix are moved to the right. This is simply to visually indicate the final matrix density. All black pixels must be covered by a sub-matrix. The coloured-shaded regions show the matrices $\{V'_x\}$ that can be calculated faster than the full matrix product.  See text for details. }
    \label{fig:matrix_visual}
\end{figure*}

A more generic idea to optimise the choice of $\{\bs{A}_x\}$ and $\{\bs{B}_x\}$ is to place down sub-matrices that cover all unique entries of $\bs{V}$, but are as small as possible in size. A heuristic algorithm for this is the following:
\begin{enumerate}[leftmargin=\parindent,align=left]
    \item Form a list of pairs of antennas, $\mathcal{U} = [(a_1, b_1), (a_2, b_2), \dots, (a_{N_u}, b_{N_u})]$ where each pair is a representative from one of the `redundant groups'. Since the choice of representative is arbitrary, there are many ways to do this, but one simple way is to use the first pair of the group in lexicographic order. 
    \item Let $\alpha = \{a_1, a_2, \dots \}$ be the set of all `first' antennas in $\mathcal{U}$, and $\beta$ the corresponding set of `second' antennas. Let $N_\alpha$ be the set of multiplicities of $a_x$ in $\mathcal{U}$. 
    \item Obtain $\alpha'$, which is a list of the elements of $\alpha$, sorted in descending order of their multiplicity. That is, sort antennas in decreasing order of how many times they appear as first-antenna in a unique pair.
    \item Decide on a number $N_{\rm sub}$ of sub-matrices into which $\bs{V}$ should be partitioned: our partition will be limited to an ordered partition of $\alpha'$, i.e. the placement of $N_{\rm sub} - 1$ cuts in $\alpha'$ such that 
    we have a new set of sub-lists, $\{\alpha'_x\}$, for $0 \leq x < N_{\rm sub}$. Let the size of each subset be $n_x$, so that $\sum_x n_x = |\alpha|$; then $\alpha'_0$ corresponds to the first $n_0$ values in $\alpha'$, $\alpha'_1$ the next $n_1$ values, etc.
    This defines the set $\{\bs{A}_x\}$ such that each element is the matrix formed by taking the rows of $\bs{Z}$ corresponding to the indices in $\alpha'_x$.
    Moreover, it uniquely defines the corresponding set $\{ \beta_x \}$ which must be the set of all $b_i$ that exist in a pair in $\mathcal{U}$ where the first element is in $\alpha'_x$. The set $\{\bs{B}_x\}$ is the analogue of $\{\bs{A}_x\}$ for $\{\beta_x\}$.
    Letting $\mathcal{V}_x$ be the outer-product of pairs from $\alpha'_x$ and $\beta'_x$, $\mathcal{V}_x = \{(i, j)\} \forall i \in \alpha'_x, \forall j \in \beta_x$,
    we denote the `size' of a particular sub-matrix as $\mathcal{S}_x = |\mathcal{V}_x|$, and its `density' $\mathcal{D}_x$ is the ratio of the number of unique pairs in $\mathcal{U}$ that appear in $\mathcal{V}_x$ to $\mathcal{S}_x$, i.e. $\mathcal{D}_x = |\mathcal{U} \cap \mathcal{V}_x| / |\mathcal{V}_x|$.
    \item To decide where the $N_{\rm sub} - 1$ partitions should be placed, use an iterative procedure. First, notice that increasing the size of a sub-matrix $\bs{A}_x$ by including one extra index from $\alpha'$ is almost always going to reduce the density $\mathcal{D}_x$, since $\alpha'$ is ordered by multiplicity. Using this, we simply do the following: for each partition point, defined by $n_x$, try all values from $n_x = 1, 2, \dots$, each time computing $\mathcal{D}_x$. Once $\mathcal{D}_x$ reduces below a threshold $\tau$, move to the next cut, $n_{x+1}$ and do the same. After placing all cuts, compute $\mathcal{D}_{N_{\rm sub}}$ for the last sub-matrix. If it is greater than $\tau$, then we're done. If not, reduce $\tau$ by some pre-defined small amount, and redo the loop over $x$ with the new threshold. 
\end{enumerate}
The output of this algorithm is the ordered list $\alpha'$ and the set of cuts, $\{n_1, n_2, \dots, n_{N_{\rm sub}}\}$, such that the densities $\mathcal{D}_x$ are as high as possible while all remaining similar. Since computation time will scale roughly as the total size, $\mathcal{S} = \sum_x \mathcal{S}_x$, we note that $\mathcal{S} = N_u / \mathcal{D}$, where by construction $\mathcal{D} < {\rm min}\{\mathcal{D}_x \}$. 
This process is illustrated in Figure \ref{fig:matrix_visual}.
This algorithm is by no means guaranteed to return the best possible decomposition of $\bf{V}$: it is constrained to partitions only in $\alpha'$, not the joint space of $\alpha'$ and $\beta$ (in the visual representation of Figure \ref{fig:matrix_visual}, different sub-matrices never appear on the same row), and it does not attempt to use other pairs from the redundant groups instead of the arbitrarily chosen representative pair. Nevertheless, it obviously does reasonably well in the case of HERA, with a compression factor of more than 15 for 3 sub-matrices compared to the full calculation.
A trivial adaptation to the method is to reverse the pairs (i.e. pass $(b, a)$ into the algorithm instead of $(a, b)$). Our tests show a compression factor of $\sim25$ in this case.

Figure \ref{fig:perf_test} shows the results of this sub-matrix method in red and purple (triangles), for 2- and 3-sub-matrices respectively. We firstly note that these two decompositions yield very similar performance. Overall they scale as we would expect: linearly with $N_{\rm ant}$ like the other methods that utilise redundancy, and linearly with $N_{\rm src}$. Since each of the sub-matrices is reasonably large, we do not see issues of under-utilisation of the GPU parallelisation for small number of sources, like we did in the VV method. 
The reason that both 2- and 3-sub-matrix cases perform similarly well seems to stem from the fact that they have similar total sizes (11.4\% and 6.6\% respectively, c.f. Fig. \ref{fig:matrix_visual}), and that the modest decrease in raw computations afforded by three sub-matrices is counter-balanced by the overhead of an extra independent matrix multiplication. It is unlikely, at least in the case of HERA, that further sub-matrices would be effective, for this reason.

The strongest downside to this method is that, similar to the MV method, it needs to copy memory on the GPU to form the $\bs{A}_x$ and $\bs{B}_x$. Our tests show that this memory copying constitutes about 1/3 of the total runtime in our non-optimised implementation. Better memory management could make this method even more competitive. 
This is seen also in the $N_{\rm side} = 512$ case, where there is a sudden jump in runtime. This is due to the extra memory consumption of this method; to get around this, a loop over chunks of $N_{\rm src}$ is required, with the extra memory copies associated. 
The sub-matrix method may also not work when baselines are not perfectly redundant since matrix compression utilise redundancies. 
We plan to improve on the method and conduct thorough tests before adding it to \matvis{} in the near future.

\section{The Square Root of the Coherency Matrix}
\label{app:mmatrix}
The form of the matrix $\boldsymbol{M}$, where
\begin{equation}
    \mathbf{M}\mathbf{M}^\dagger = \boldsymbol{\mathsf{C}} = \begin{pmatrix}
        I + Q & U + iV \\
        U - iV & I - Q
    \end{pmatrix},
\end{equation}
is 
\begin{equation}
    \mathbf{M} = \frac{1}{\sqrt{2}}\begin{pmatrix}
        \frac{(Q-T)\sqrt{\frac{IQ}{T} + I - T - Q}}{(U-iV)} & \frac{(T +Q)\sqrt{-\frac{IQ}{T}+I+T-Q}}{(U-iV)} \\
        \sqrt{(Q/T + 1)(I - T)} & \sqrt{(1 - Q/T)(I + T)}
    \end{pmatrix},
\end{equation}
where $T^2 = Q^2 + U^2 + V^2$.

\section{Author Affiliations}\label{sec:affiliations}
\noindent\textit{$^{1}$ Department of Physics and Astronomy,  University of Western Cape, Cape Town, 7535, South Africa\\
$^{2}$ South African Radio Astronomy Observatory, Black River Park, 2 Fir Street, Observatory, Cape Town, 7925, South Africa\\
$^{3}$ Scuola Normale Superiore, 56126 Pisa, PI, Italy\\
$^{4}$ School of Earth and Space Exploration, Arizona State University, Tempe, AZ\\
$^{5}$ Queen Mary University London, London E1 4NS, UK\\
$^{6}$ Jodrell Bank Centre for Astrophysics, University of Manchester, Manchester, M13 9PL, United Kingdom\\
$^{7}$ Department of Astronomy, University of California, Berkeley, CA\\
$^{8}$ Department of Physics and Astronomy, University of Pennsylvania, Philadelphia, PA\\
$^{9}$ Department of Physics, Winona State University, Winona, MN\\
$^{10}$ Department of Physics and Trottier Space Institute, McGill University, 3600 University Street, Montreal, QC H3A 2T8, Canada\\
$^{11}$ INAF-Istituto di Radioastronomia, via Gobetti 101, 40129 Bologna, Italy\\
$^{12}$ Department of Physics and Electronics, Rhodes University, PO Box 94, Grahamstown, 6140, South Africa\\
$^{13}$ National Radio Astronomy Observatory, Charlottesville, VA\\
$^{14}$ Cavendish Astrophysics, University of Cambridge, Cambridge, UK\\
$^{15}$ National Radio Astronomy Observatory, Socorro, NM 87801, USA\\
$^{16}$ MIT Kavli Institute, Massachusetts Institute of Technology, Cambridge, MA\\
$^{17}$ Department of Physics, Massachusetts Institute of Technology, Cambridge, MA\\
$^{18}$ Centre for Strings, Gravitation and Cosmology, Department of Physics, Indian Institute of Technology Madras, Chennai 600036, India\\
$^{19}$ Radio Astronomy Lab, University of California, Berkeley, CA\\
$^{20}$ Department of Physics, University of California, Berkeley, CA\\
$^{21}$ Department of Physics and Astronomy, University of California, Los Angeles, CA\\
$^{22}$ National Radio Astronomy Observatory, Socorro, NM\\
$^{23}$ Institut d’Astrophysique Spatiale, CNRS, Université Paris-Saclay, 91405 Orsay, France\\
$^{24}$ School of Physics, University of Melbourne, Parkville, VIC 3010, Australia\\
$^{25}$ Department of Physics, University of Washington, Seattle, WA\\
$^{26}$ eScience Institute, University of Washington, Seattle, WA\\
$^{\dagger}$ NASA Hubble Fellow\\
$^{27}$ Department of Physics, Brown University, Providence, RI\\
$^{28}$ UNLV\\
$^{29}$ Department of Physics and McGill Space Institute, McGill University, 3600 University Street, Montreal, QC H3A 2T8, Canada\\
$^{30}$ Department of Physics, Stellenbosch University, Matieland, Western Cape, 7602, South Africa\\
$^{31}$ School of Chemistry and Physics, University of KwaZulu-Natal, Westville Campus Private Bag X54001, Durban, 4000, South Africa\\
$^{32}$ International Centre for Radio Astronomy Research, Curtin University, Bentley, WA 6102, Australia\\
$^{33}$ ARC Centre of Excellence for All Sky Astrophysics in 3 Dimensions (ASTRO 3D), Bentley, WA 6102, Australia\\
$^{34}$ School of Physics, University of Melbourne, Parkville, VIC 3010 Australia\\
$^{35}$ Raman Research Institute\\
$^{36}$ Commonwealth Scientific and Industrial Research Organisation (CSIRO), Space \& Astronomy, P. O. Box 1130, Bentley, WA 6102, Australia
}

% Don't change these lines
\bsp	% typesetting comment
\label{lastpage}
\end{document}